\documentclass[fleqn,usenatbib]{mnras}

\usepackage{newtxtext,newtxmath}

\usepackage[T1]{fontenc}
\usepackage{ae,aecompl}

\usepackage{graphicx}	
\usepackage{amsmath}	
\usepackage{braket}

\title[]{Modelling of the multi-transition periodic flaring in G9.62+0.20E}

\author[F. Rajabi et al.]{F. Rajabi,$^{1,2}$\thanks{E-mail: fereshte.rajabi@gmail.com} M. Houde,$^{1}$ G. C. MacLeod,$^{3,4}$ S. Goedhart,$^{5,6}$ Y. Tanabe,$^{7}$ 
\newauthor S. P. van den Heever,$^{4}$ C. M. Wyenberg,$^{8}$ Y. Yonekura$^{7}$  
\\
$^{1}$Department of Physics and Astronomy, The University of Western Ontario, 1151 Richmond Street, London, Ontario N6A 3K7, Canada \\
$^{2}$Perimeter Institute for Theoretical Physics, Waterloo, Ontario N2L 2Y5, Canada \\
$^{3}$The Open University of Tanzania, P.O. Box 23409, Dar-Es-Salaam, Tanzania \\
$^{4}$Hartebeesthoek Radio Astronomy Observatory, PO Box 443, Krugersdorp, 1741, South Africa \\
$^{5}$South African Radio Astronomy Observatory, 2 Fir Street, Black River Park, Observatory 7925, South Africa \\
$^{6}$Center for Space Research, North-West University, Potchefstroom Campus, Private Bag X6001, Potchefstroom 2520, South Africa \\
$^{7}$Center for Astronomy, Ibaraki University, 2-1-1 Bunkyo, Mito, Ibaraki 310-8512, Japan \\
$^{8}$Institute for Quantum Computing and Department of Physics and Astronomy, The University of Waterloo, 200 University Ave. West, Waterloo, \\ 
$^{\;}$Ontario N2L 3G1, Canada 
}

\date{}

\begin{document}
\label{firstpage}
\pagerange{\pageref{firstpage}--\pageref{lastpage}}
\maketitle

\begin{abstract}
We present detailed modeling of periodic flaring events in the 6.7~GHz and 12.2~GHz methanol lines as well as the OH~1665~MHz and 1667~MHz transitions observed in the G9.62+0.20E star-forming region. Our analysis is performed within the framework of the one-dimensional Maxwell-Bloch equations, which intrinsically cover the complementary quasi-steady state maser and transient superradiance regimes. We find that the variations in flaring time-scales measured for the different species/transitions, and sometimes even for a single spectral line, are manifestations of and are best modeled with Dicke's superradiance, which naturally accounts for a modulation in the duration of flares through corresponding changes in the inversion pump. In particular, it can explain the peculiar behaviour observed for some features, such as the previously published result for the OH~1667~MHz transition at $v_\mathrm{lsr}=+1.7$~km~s$^{-1}$ as well as the methanol 6.7~GHz line at $v_\mathrm{lsr}=-1.8$~km~s$^{-1}$, through a partial quenching of the population inversion during flaring events.
\end{abstract}




\section{Introduction}\label{sec:Introduction}

G9.62+0.20E is a star-forming region located at a distance of 5.2~kpc from the Sun \citep{Sanna2009} and home to a number of young massive stars going through the early stages of their evolution \citep{Hofner1996}. \citet{Goedhart2003} reported the monitoring in this source in the methanol 6.7~GHz, starting in 1999, and 12.2~GHz, from 2000, maser lines and discovered six flares igniting periodically every $\sim 246$ days. Each flare lasted approximately three months and peaked within a month. \citet{Caswell1998} had also previously reported the detection of OH and water masers in this region, while high resolution imaging of some of the strongest 6.7~GHz methanol maser in this source were associated to ionized stellar winds around a young massive star \citep{Hofner1996}. 

\citet{VanderWalt2009} summarized the results of their 2600-day monitoring of 6.7~GHz and 12.2~GHz methanol masers in G9.62+0.20E and found a similar periodicity (244 days) for the 6.7~GHz transition flares. They also determined that the flux densities of 6.7~GHz and 12.2~GHz lines returned to their same respective levels during the low intensity phase of the cycles. This led to the conclusion that the source of periodicity must be outside the masing region. Later on, \citet{Goedhart2019} reported the results of their simultaneous monitoring of the OH main lines (i.e., at 1665 and 1667~MHz), the 6.7~GHz and 12.2~GHz methanol and the 22~GHz water transitions. While the OH main lines were monitored using KAT-7 (seven-element Karoo Array Telescope), the methanol and water data were obtained using the Haartebeesthoek Radio Astronomy Observatory (HartRAO) 26-m telescope. They observed simultaneous flaring activity in all those lines for some velocity components. One notable feature of their observations is the apparent delay between the flares of the different species/transitions. More precisely, while the methanol lines at some velocities start to flare when OH transitions exhibit a marked drop in their intensity, the delay between the peak intensity of the 22~GHz water is most pronounced relative to those of methanol. Using previous observations from \citet{Sanna2009}, \citet{Goedhart2019} also determined that the sources of methanol and OH flares were located some 1600~AU apart. Using this positional arrangement they put forth a model based on changes in the seed radiation to possibly explain the differences in the flaring between the OH and 12.2~GHz methanol lines. However since the OH main lines and the methanol 12.2~GHz (as well as 6.7~GHz) lines are radiatively pumped  \citep{Gray2012}, a change in the intensity of the pump would thus also be expected to translate in changes in the lines' flux densities. The emergence of peak intensity at one line when the other also radiatively pumped line goes quiet is a nontrivial behaviour that warrants a more careful analysis.

Recently, the underlying nature of G9.62+0.20E has revealed itself to be even more complex than believed thus far. Using observations of methanol at 6.7~GHz from the Hitachi 32~m telescope of the Ibaraki station from the NAOJ Mizusawa VLBI Observatory, \citet{MacLeod2022} uncovered a second period of 52~d at $v_{\mathrm{lsr}} = +8.8$~km s$^{-1}$ in this source. Given that the two known periods (i.e., 243~d and 52~d) are not harmonically related, it is likely that they are produced by two independently varying excitation sources located within G9.62+0.20E.   

Simultaneous flaring but with some delay between the flares of different lines or species is not unique to the observations of \citet{Goedhart2019} in G9.62+0.20E. \citet{MacLeod2019} reported the detection of several new methanol transitions flaring on similar time-scales as the 6.7~GHz line of the same species in G358.93--0.03, while \citet{MacLeod2021} confirmed the periodic flaring in multiple OH (including excited-OH at 6.031~GHz and 6.035~GHz) and methanol transitions in G323.459--0.079. \citet{Araya2010} found quasi-periodic and simultaneous flaring in the 6~cm formaldehyde and 6.7~GHz methanol spectral lines in IRAS 18566+0408, although the flares were not spatially coincident. \citet{Seidu2022} also reported periodic flaring of OH main lines and the 6.7~GHz methanol line for some velocity components in G339.62--0.12. The peak intensities of different lines appeared with some time delays on the order of a few to 15 days, while they also noted that the methanol and OH masers were not spatially coincident either. \citet{Szymczak2016} also reported observations of alternating 6.7~GHz methanol and 22~GHz water flares which appeared periodically in G107.298+5.639. The delay between the peak of water flares and that of the methanol flares in some velocity components was about 18 days. An interesting aspect of this discovery was that all the periodic water masers originated from the 360~au methanol maser region. Since the 22~GHz water masers are mainly pumped through collisions and the densities and temperatures required for the pumping of 6.7~GHz and the 22~GHz water do not intersect broadly, the discovery of coincident 6.7~GHz methanol and 22~GHz water in some velocity components is of significant importance in understanding these types of flaring activities. \citet{Rajabi2017} used Dicke's superradiance model to explain the alternating methanol and water flares in G107.298+5.639 and provided an alternative picture to that based on maser theory. 

Dicke's superradiance is in essence a cooperative spontaneous emission process, through which radiators (excited atoms/molecules) release their energy in a much more efficient way \citep{Dicke1954}. More precisely, a group of $N$ excited atoms/molecules couple to one another through interactions with their common electromagnetic field and act as a single macro-radiator with a cooperative time-scale scaling as $\left(N\Gamma\right)^{-1}$, with $\Gamma$ is the spontaneous emission rate from a single atom/molecule.

In \citet{Rajabi2016A, Rajabi2016B, Rajabi2017, Rajabi2019}, signatures of Dicke's superradiance in maser harboring regions and mainly in star-forming regions were presented at the 6.7~GHz methanol, 22~GHz water and 1612~MHz OH lines. Common features among flares modelled with superradiance are a sharp rise in the flux density of the line, sometimes at a faster rate than the corresponding pump, as well as measurable variations in time-scales (delays and duration) between different spectral transitions. In short, Dicke's superradiance highlights a fast transient regime of radiation that is complementary to the maser action, which is relatively speaking a slow and quasi-steady state process. In \citet{Rajabi2020}, it is shown how a system can transition between the maser and superradiance regimes when certain criteria are met. 

This paper is structured as follows. In Sec. \ref{sec:SRvsMaser}, the complementarity of the maser and superradiance regimes is further discussed with more details on observational features distinguishing the two phenomena. After a brief review of some of the existing theoretical models for the flaring activities reported in G9.62+0.20E and a description of the data used for our analyses in Secs. \ref{sec:TheoryReview}--\ref{sec:data} we focus, in Sec. \ref{sec:results}, on the analysis and modeling of these data within the framework of the one-dimensional Maxwell-Bloch equations, which intrinsically cover the complementary quasi-steady state maser and transient superradiance regimes. Models for periodic flares at 6.7~GHz methanol, 12.2~GHz methanol and 1665/1667~MHz OH lines in G9.62+0.20E are shown and the characteristics of the underlying systems presented. We end with a discussion of our results and broader implications for maser-hosting systems in Sec. \ref{sec:discussion}. 

\section{The maser and superradiance regimes}\label{sec:SRvsMaser}  
The evolution of a compound system composed of an extended group of molecules (or atoms) and a radiation field is described through the Maxwell-Bloch equations (MBE), which track the underlying difference in population levels, the polarisation (for electric dipole transitions) and the radiation field \citep{MacGillivray1976,Gross1982,Benedict1996}. A close study of the MBE for a gas initially hosting a population inversion reveals two limiting and complementary regimes: the quasi-steady state regime where the maser action takes place (and the stimulated emission process dominates) and the fast transient regime where superradiance is at play \citep{Feld1980,Rajabi2020}. 

\begin{figure}
    \centering
    \includegraphics[width=0.47\textwidth]{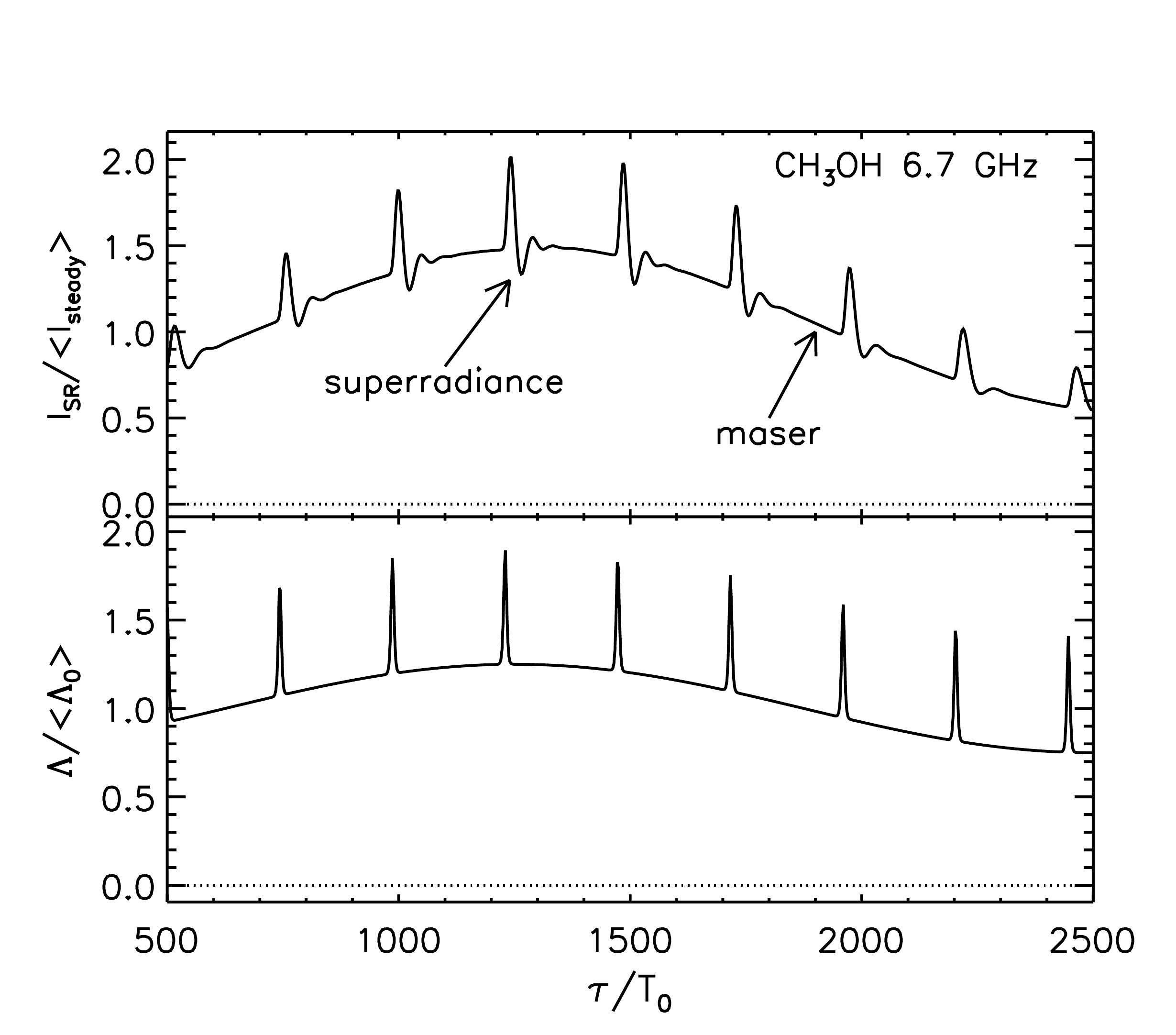}
    \caption{Mean-normalized output intensity of a radiating system (top graph) in response to an inversion pump (bottom graph). The pump consists of a slow component oscillating on a time-scale $\sim 1000\,T_0$ about a mean value $\left<\Lambda_0\right>$ and a fast excitation varying periodically on a much shorter time scale ($\lesssim 10\,T_0$with a period of $243.3\,T_0$; $T_0=1\,\mathrm{day}$). The radiation system, modelled for the methanol 6.7~GHz spectral line, exhibits a fast transient superradiance response that features a damped oscillating behaviour and is broader than the excitation signal from the fast periodic pump component. The superradiance transient is superposed on a quasi-steady state response in the saturated maser regime that closely tracks the slowly oscillating pump component. For this example we set $T_1=220\,T_0$ and $T_2=12.6\,T_0$.}
    \label{fig:steady+transient}
\end{figure}

In the quasi-steady state maser regime temporal variations in the population inversion, polarization and radiation field happen on an evolution time-scale $T_\mathrm{e}$ longer than the decaying and dephasing time-scales $T_1$ and $T_2$ characterizing the response of the system (e.g., $\partial P/\partial\tau\sim P/T_\mathrm{e} \ll P/T_2$ with $P$ the polarization; see equations \ref{eq:dN/dt}-\ref{eq:dE/dz} below and \citealt{Rajabi2020}). As a result of this a maser system will quickly adjust to and can follow the excitation signal. For example, the output intensity of a saturated maser system will tend to mimic the temporal behaviour of the inversion pump, while that of an unsaturated maser will follow the seed signal. An example of this is shown in Figure \ref{fig:steady+transient} for a methanol 6.7~GHz radiating system (top graph) subjected to an inversion pump (bottom graph). The pump consists of a slow component oscillating on a time-scale $\sim 1000\,T_0$ about a mean value $\left<\Lambda_0\right>$ and a fast excitation varying periodically on a much shorter time scale ($\lesssim 10\,T_0$ with a period of $243.3\,T_0$; $T_0=1\,\mathrm{day}$). The system exhibits a fast transient superradiance response featuring a damped oscillating behaviour that is broader than the excitation signal from the fast periodic pump component. The superradiance signal is superposed on a quasi-steady state response in the saturated maser regime that closely tracks the slowly oscillating pump component. Contrary to the superradiance regime the saturated maser response of the system does not display any obvious transient to the slow varying pump stimulus but quickly adjusts to any changes in it. We note that the presence of two widely differing periods has been observed in maser-hosting systems (see \citealt{Tanabe2023} for periods of $130.6$~d and $\sim1260$~d in G5.900--0.430).  

The system's singular behaviour in the fast transient superradiance regime rests on the fact that it responds in a totally different manner to a variation in the pump signal whenever $T_\mathrm{e}\sim T_\mathrm{R}\ll T_1,T_2$, with $T_\mathrm{R}$ the time-scale of superradiance (defined in equation \ref{eq:TR} below). As will soon be shown the transient intensity triggered by the excitation is a characteristic response of the system in the superradiance regime and is largely independent of the shape of the input signal. Another salient feature of the superradiance regime is the fact that the peak intensity scales as $N^2$, with $N$ the number of molecules involved, as opposed to a scaling with $N$ for a saturated maser \citep{Rajabi2020}. It is important to note that the details of the transient response (e.g., its time-scale, shape and amplitude) depend on the physical conditions characterizing the environment where the atomic/molecular population is found (e.g., through the importance of the time-scales $T_1$ and $T_2$ relative to $T_\mathrm{e}$, and the size of the column density) and the parameters characterizing the radiative transition under study (i.e., its wavelength and Einstein spontaneous emission coefficient; see below).

\begin{figure*}
    \centering
    \begin{minipage}[t]{.48\textwidth}
    \includegraphics[width=\columnwidth]{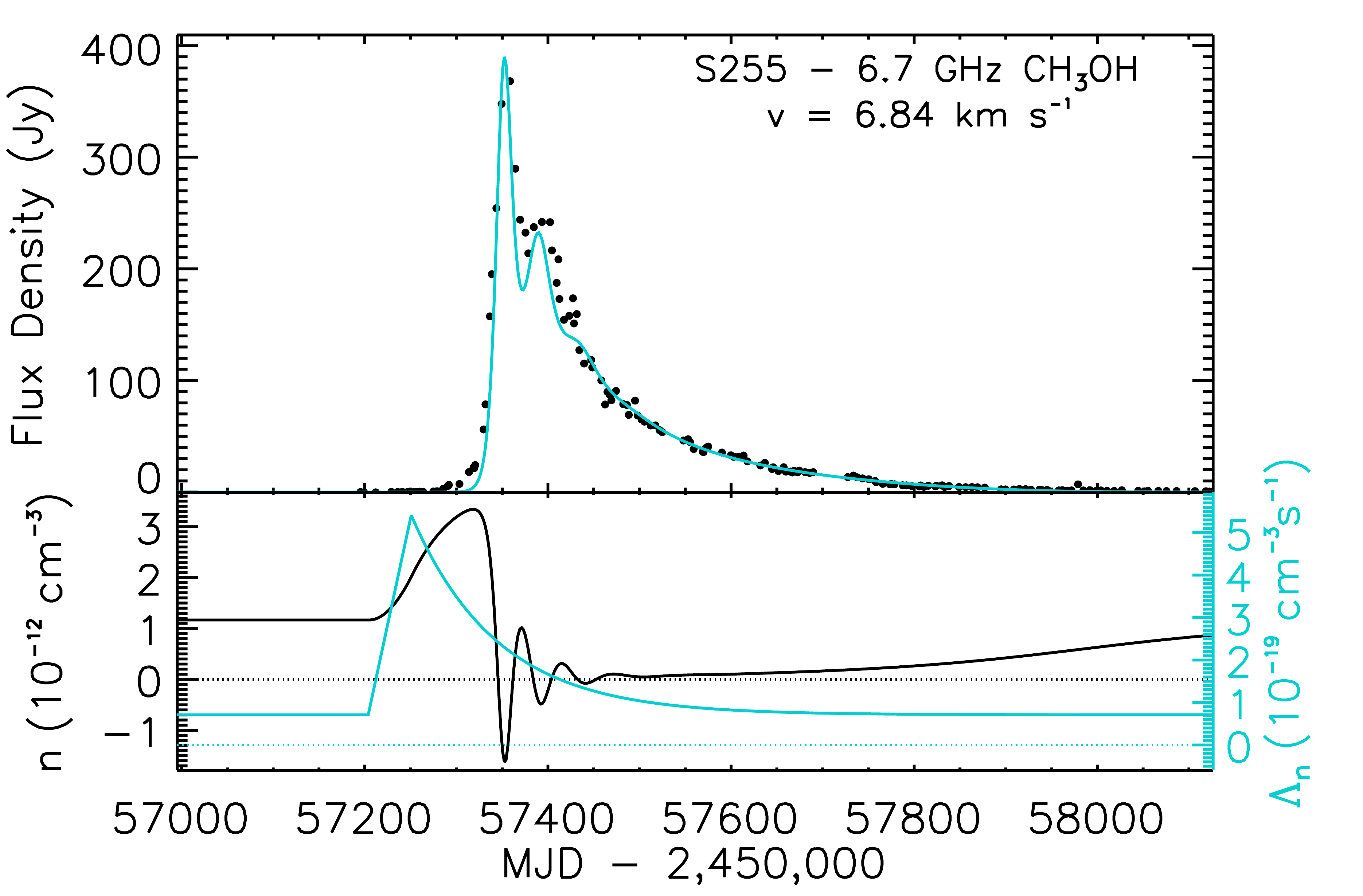}
    \end{minipage}\qquad
    \begin{minipage}[t]{.48\textwidth}
    \includegraphics[width=\columnwidth]{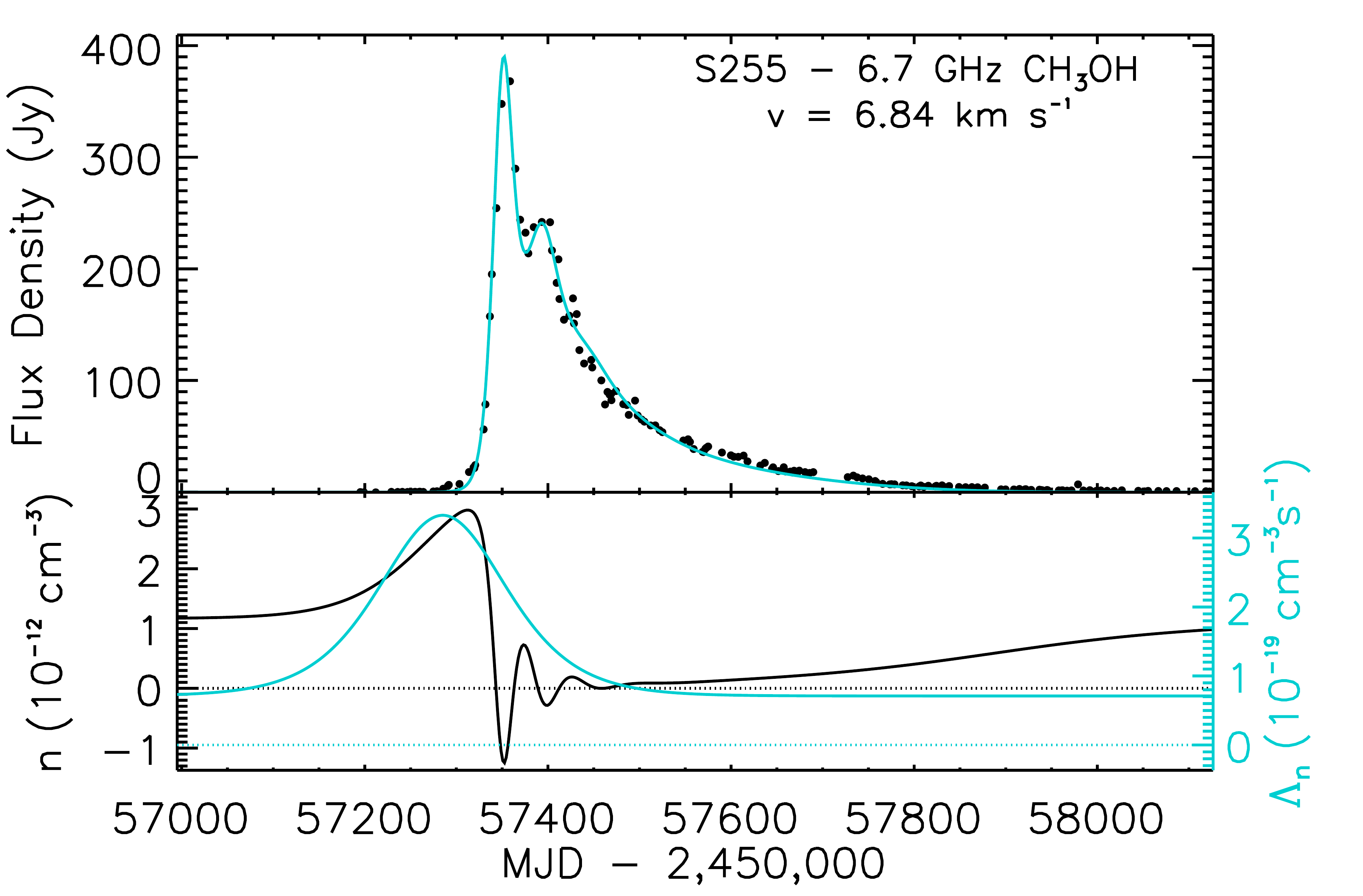}
    \end{minipage}
    \caption{Superradiance model for the S255IR-NIRS3 6.7~GHz methanol flare at $v_{\mathrm{lsr}} = 6.84~\mathrm{km~s}^{-1}$ \citep{Szymczak2018b,Rajabi2019,Rajabi2020}. Left panel: the top panel shows the model fit (light/cyan curve) to the data (dots), while the bottom panel is for the temporal evolution for the pump (light/cyan curve, vertical axis on the right) and the population inversion density (dark/black curve, vertical axis on the left). We used a slow rising exponential function for the pump pulse to approximately mimic the IR observations presented in \citet{Uchiyama2020}. Right panel: same as the left panel, but with a smoother and symmetric function for the pump signal (adapted from \citealt{Rajabi2020}). The parameters for both models are $L=2.17\times 10^{15}$~cm, $T_1=1.64 \times 10^{7}$~s and $T_2 =1.55 \times 10^6$~s for the decay and dephasing time-scales, respectively, while the inversion level prior to the appearance of the pump pulse corresponds to approximately $0.1~\mathrm{cm}^{-3}$ for a molecular population spanning a velocity range of $1~\mathrm{km~s}^{-1}$. The superradiance flux density is scaled to the data. These examples show that the fast transient superradiance response of the system is largely independent of the detailed nature of the inversion pump.}
    \label{fig:S255}
\end{figure*}

This difference in responses to an excitation provides us with a diagnostic tool that can be used to determine through observations when an astronomical source is in the quasi steady-state maser or the superradiance regimes. That is, the comparison of a well-sampled intensity curve of the radiation emanating at the maser transition with another light curve providing, for example, the temporal evolution of the pump signal (presumably at a much shorter wavelength) will reveal whether the shape of the former tracks (in the saturated maser regime) or not (superradiance regime) that of the latter. Evidently, this requires complementary observations at distinct wavelength ranges that are not always easily obtainable and, accordingly, few cases where this was achieved exist. One notable example is the monitoring of the S255IR-NIRS3 at the 6.7~GHz methanol transition \citep{Szymczak2018b} and the contemporaneous $Ks$ band IR light curve \citep{Uchiyama2020}. In this case the shape of the slow rising exponential looking IR curve is not matched by that of the 6.7~GHz methanol intensity at any velocity (see Fig. 3 in \citealt{Uchiyama2020}), while the increase in flux densities at the maser transition in some cases greatly exceed that in the IR (which increases by a factor of approximately 20; see below). The fact that these observations point to S255IR-NIRS3 being in the fast transient regime is supported by the possibility of modelling the 6.7~GHz methanol flux density at some velocities with a superradiance response \citep{Rajabi2019,Rajabi2020}. We show in the left panel of Figure \ref{fig:S255} an example where we used a slow rising exponential function for the pump pulse (bottom graph, cyan curve) to approximately mimic the IR observations presented in \citet{Uchiyama2020} and fit the data of \citet{Szymczak2018b} at $v_\mathrm{lsr}=6.84\,\mathrm{km\,s^{-1}}$ using the same parameters as in \citet{Rajabi2020} (top graph; see also \citealt{Rajabi2019}). The aforementioned near-independence of the shape and amplitude of the flux density curve from that of inversion pump can be surmised in the right panel of Figure \ref{fig:S255} where the slow-rising exponential was replaced by a smoother and symmetric function for the pump signal. Despite the obvious differences between the inversion pump signals the responses of the system remains practically unchanged. 

The multi-transition monitoring of periodically flaring  sources, such as G9.62+0.20E, presents us with another opportunity to distinguish between the maser and superradiance regimes, which does not necessitate information about the excitation signal (e.g., the pump signal in the preceding discussion). As was already mentioned, in the quasi steady-state regime corresponding to the saturated maser the shape of the excitation signal is passed on in the response of the system, as observed in the intensity of the source. For example, the time-scale in variability will be set by the excitation and appear in a similar manner in different spectral transitions with which the source might be detected. 

On the other hand, for superradiance the time-scale for the evolution of the system is given by \citep{MacGillivray1976,Gross1982,Benedict1996,Rajabi2020}
\begin{align}
    T_\mathrm{R} = \frac{8\pi}{3nL\lambda^2}\tau_\mathrm{sp},\label{eq:TR}
\end{align}
where $\tau_\mathrm{sp}=\Gamma^{-1}$ is the spontaneous emission time-scale of the transition considered, $\lambda$ the wavelength of radiation, $n$ the population inversion level and $L$ the length of the system ($nL$ is therefore the column density).\footnote{We are considering radiation systems of cylindrical shape.} It follows from equation (\ref{eq:TR}) that the speed at which a superradiance system responds to an excitation (i.e., the transient response) will vary from one transition to another \citep{Rajabi2017}. For example, the ratio of the time-scales from the 6.7~GHz methanol and 1665~MHz OH spectral lines is (from equation \ref{eq:TR}) 
\begin{align}
    \frac{T_\mathrm{R,1665\,MHz}}{T_\mathrm{R,6.7\,GHz}} & = \frac{\left(nL\right)_\mathrm{6.7\,GHz}}{\left(nL\right)_\mathrm{1665\,MHz}}\frac{\Gamma_\mathrm{6.7\,GHz}}{\Gamma_\mathrm{1665\,MHz}}\left(\frac{1.665\,\mathrm{GHz}}{6.7\,\mathrm{GHz}}\right)^2\\
    & \simeq 1.4\frac{\left(nL\right)_\mathrm{6.7\,GHz}}{\left(nL\right)_\mathrm{1665\,MHz}},\label{eq:TR-ratio-1665_6.7}
\end{align}
which implies that for equal column densities the 1665~MHz OH transition will be approximately 40\% slower than the 6.7~GHz methanol line. Other time-scale ratios that will be relevant for our analysis involve the 1667~MHz OH and 12.2~GHz methanol transitions
\begin{align}
    \frac{T_\mathrm{R,1667\,MHz}}{T_\mathrm{R,6.7\,GHz}} & \simeq 1.3\frac{\left(nL\right)_\mathrm{6.7\,GHz}}{\left(nL\right)_\mathrm{1667\,MHz}}\label{eq:TR-ratio-1667_6.7}\\
    \frac{T_\mathrm{R,12.2\,GHz}}{T_\mathrm{R,6.7\,GHz}} & \simeq 0.6\frac{\left(nL\right)_\mathrm{6.7\,GHz}}{\left(nL\right)_\mathrm{12.2\,GHz}}.\label{eq:TR-ratio-12.2_6.7}
\end{align}
 
It therefore follows that we should expect that fast transient responses from different spectral transitions will happen on varying time-scales when they are subjected to the same excitation (i.e., either in the population inversion pump or the seed radiation). 

\begin{figure}
    \centering
    \includegraphics[width=0.47\textwidth]{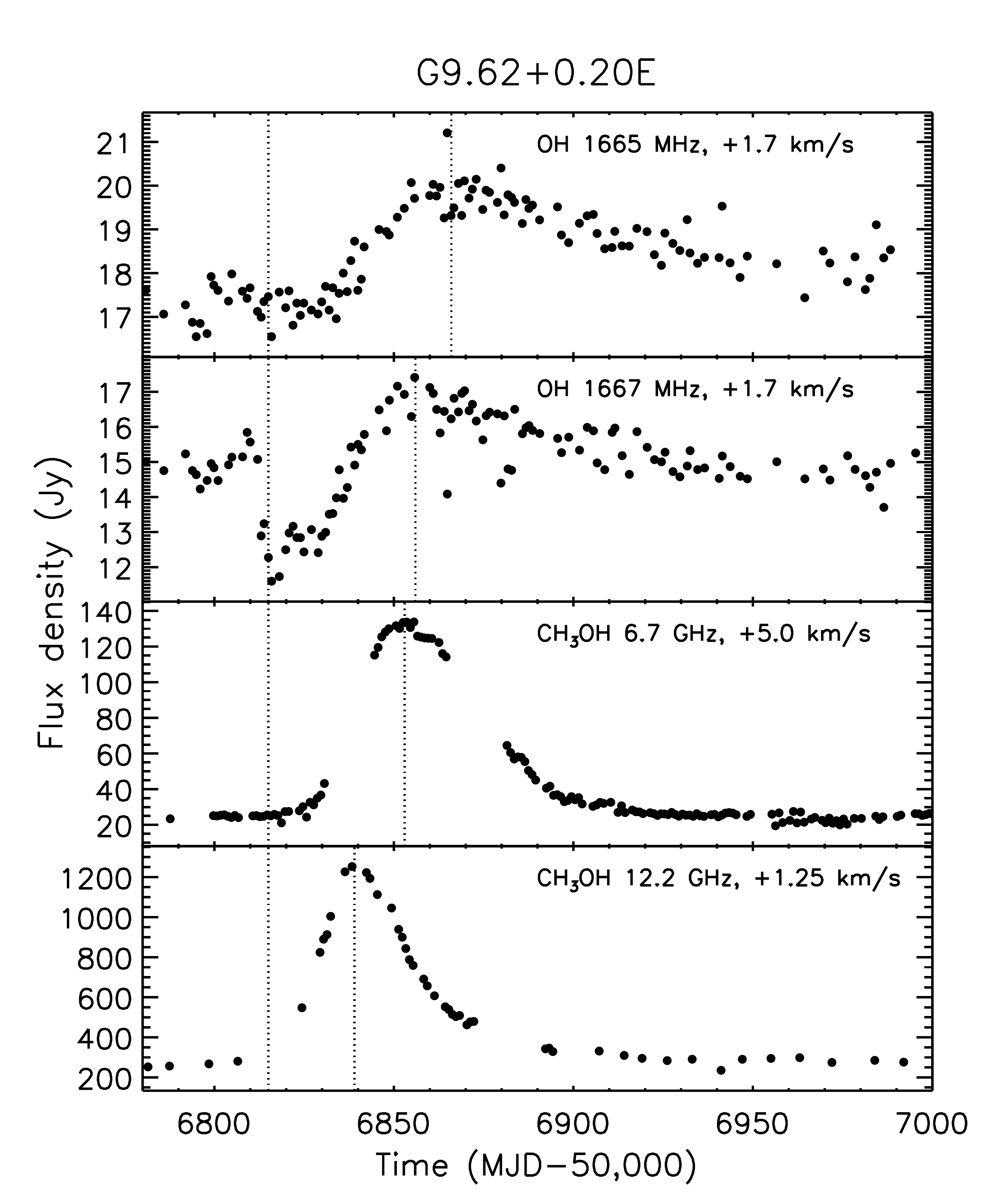}
    \caption{Comparison of profiles between OH and methanol flares, similar to Fig. 8 of \citet{Goedhart2019} but with the addition of the methanol 6.7~GHz feature. The vertical line at MJD 56815 denotes the start of the methanol 12.2~GHz flare, as defined by \citet{Goedhart2019}. The other vertical lines at later times indicate the approximate occurrence of maximum intensity for each flare. The methanol 6.7~GHz light curve was delayed by 15 days to better align with the three others.}
    \label{fig:g9.62-all}
\end{figure}

Interestingly, a comparison of OH and methanol flare profiles in G9.62+0.20E provides some indication of varying time-scales for the different spectral transitions. This is shown in Figure \ref{fig:g9.62-all} where the OH 1665 and 1667~MHz at $+1.7$~km~s$^{-1}$ and methanol 6.7 and 12.2~GHz at, respectively, $+5.0$ and $+1.25$~km~s$^{-1}$ are presented for the flares happening at around MJD 56850. In the figure, the vertical line at MJD 56815 denotes the start of the methanol 12.2~GHz flare, as defined by \citet{Goedhart2019}, while the methanol 6.7~GHz light curve was delayed by 15 days to better align with the three others. The other vertical lines at later times indicate the approximate occurrence of maximum intensity for each flare.  Given the dependency on the ratio of column densities in equations (\ref{eq:TR-ratio-1665_6.7})-(\ref{eq:TR-ratio-12.2_6.7}), we should be careful not to put too much weight on the differences in time-scales observed in the figure. Still, it is intriguing that the four time-scales are qualitatively ordered in a manner consistent with superradiance at comparable column densities. 

\section{Existing theoretical models for multi-transitional periodic flaring in G9.62+0.20E}\label{sec:TheoryReview}

Periodic flaring is not unique to G9.62+0.20E. There has so far been a total of 29 cases exhibiting periodic flaring just at the 6.7 GHz methanol line (e.g., \citealt{Szymczak2011,Fujisawa2014c,Maswanganye2015,Szymczak2015, Maswanganye2016,Szymczak2016, Sugiyama2017,Olech2022,Tanabe2023}). The theoretical models proposed to explain periodicity in most flares evolve around periodic changes either in the corresponding excitation pump or the ability of radiating molecules/atoms to get coupled to the pump (i.e., the efficiency of the pump). For instance, for the radiatively pumped methanol 6.7~GHz line, periodic changes in the infrared pump radiation flux is proposed as an explanation for periodic flares \citep{Araya2010, Parfenov2014}. A periodic change in local dust temperature can also impact the pumping efficiency periodically and explain some of the reported flaring observations. \citet{Inayoshi2013}, for instance, proposed that some massive stars undergoing accretion can exhibit pulsation that can change the local dust temperature repeatedly. Another hypothesis evolves around colliding wind binaries (CWBs), which generate ionizing radiation waves as a result of periodic wind interactions \citep{VanderWalt2011,VanderWalt2016}. This can effectively change the local dust temperature and subsequently the flare profile. One interesting feature of this model is its ability to produce asymmetric flare profiles. However this model requires the presence of an ultracompact HII region.

As mentioned earlier, G9.62+0.20E exhibits intriguing dips in the flux density of OH main lines. This poses an extra challenge for theoretical models to explain. \citet{Goedhart2019} provided a qualitative explanation for the OH main line flux dips within the CWB picture. In short, with the CWB model changes in the flares are associated with changes in the background free-free emission from a nearby HII region. A decrease in the background free-free emission, which is thought to be igniting the OH main lines, can then be used to explain a dip in the corresponding flare profiles. \citet{Goedhart2019} propose a scenario for G9.62+0.20E in which high energy ionizing photons from the hot shocked gas move past the ionization front and create high temperature electrons through ionizing hydrogen. High temperature electrons in plasma have lower emissivity, which translate to lower free-free background emission. This in turn can result in a drop in maser intensity. We will consider an alternative to this scenario in Sec. \ref{sec:unusual}.

Finally, a recent study by \citet{Morgan2021} notes the effect of source geometry and its inclination in the plane of the sky on the observed flare profiles of the 6.7 GHz methanol line in G9.62+0.20E, as well as for G22.357+0.066 and G25.411+0.105. 

\section{Data}\label{sec:data}

The data sets used for the analyses detailed in Sec. \ref{sec:results} were taken from earlier works. More precisely, the OH~1665~MHz and 1667~MHz data modeled in Secs. \ref{sec:1665MHz} and \ref{sec:unusual}, respectively, were obtained with KAT-7 \citep{Foley2016} and were first presented in \citet{Goedhart2019}. The single-dish methanol 12.2~GHz observations were realized with HartRAO 26-m telescope and also initially published by \citet{Goedhart2019}. These data sets are the same as those mentioned in Sec. \ref{sec:Introduction} when reporting on the earlier study of \citet{Goedhart2019} on G9.62+0.20E.

The methanol 6.7~GHz observations dealt with in Secs. \ref{sec:6.7GHz} and \ref{sec:unusual} are taken from those previously presented in \citet{MacLeod2022}, which led to the discovery of the 52~d flaring period discussed in Sec. \ref{sec:Introduction}. As stated then, these data result from the corresponding monitoring of G9.62+0.20E performed with the Hitachi 32~m telescope of the Ibaraki station from the NAOJ Mizusawa VLBI Observatory \citep{Yonekura2016}.

More details on these data sets and the underlying observations will be found in the cited references, where they were first published. 

\section{Model and results}\label{sec:results}

In this section we presents models based on the one-dimensional MBE for flares observed in the spectral transitions presented in Figure \ref{fig:g9.62-all}. The models are periodic at 243.3~days with an excitation provided by the inversion pump signal. Throughout, it is assumed that the population inversion and its changes are realized through radiative pumping. Except for one case in Sec. \ref{sec:unusual}, the pumping source for the population inversion at position $z$ and retarded time $\tau=t-z/c$ is of the form 
\begin{equation}
\Lambda_{n}\left(z,\tau\right) = \Lambda_{\mathrm{0}} + \sum_{m=0}^\infty \frac{\Lambda_{\mathrm{1},m}}{\cosh^{2}\left[\left(\tau-\tau_0-m\tau_1\right)/T_\mathrm{p}\right]},\label{eq:pump}
\end{equation}
and is propagating along the symmetry axis of the system (i.e., the $z$-axis). A constant pump rate $\Lambda_{\mathrm{0}}$ is applied, while $\Lambda_{\mathrm{1},m}$ is the amplitude of pump pulse $m$ at $\tau=\tau_0+m\tau_1$, with $\tau_0$ a delay adjusted to line our model up to the corresponding data and $\tau_1=243.3$~days. All of our fits contain at least two intensity bursts, and the pump pulse amplitude $\Lambda_{\mathrm{1},m}$ is adjusted to match that of the bursts. The width of the pump pulse is set by the time-scale $T_{\mathrm{p}}$, which we fix to $4$~days for all of our models. The details and choice of the pump pulse's profile are arbitrary, as long as it is shorter than the flares' duration.    

Otherwise, we proceed in a manner similar to \citet{Rajabi2019} and \citet{Rajabi2020b} in that we solve the MBE for a two-level system at resonance using a fourth-order Runge-Kutta method with the slow-varying-envelope and rotating wave approximations \citep{Mathews2017,Houde2018b,Rajabi2019,Rajabi2020}. These equations are 
\begin{align}
     & \frac{\partial n^\prime}{\partial\tau} = \frac{i}{\hbar}\left(P^+E^+-E^-P^-\right)-\frac{n^\prime}{T_1}+ \Lambda_n \label{eq:dN/dt} \\
     & \frac{\partial P^+}{\partial\tau} = \frac{2id^2}{\hbar}E^-n^\prime-\frac{P^+}{T_2}+\Lambda_P \label{eq:dP/dt} \\
     & \frac{\partial E^+}{\partial z} = \frac{i\omega_0}{2\epsilon_0c} P^-\label{eq:dE/dz},
\end{align}
where $n^\prime$ is (half of) the inverted population density, while $P^+$ and $E^+$ are the amplitudes of the molecular polarisation and the electric field, respectively. The superscript ``$+$'' is for the polarization corresponding with the molecular transition from the lower to the upper level and the positive frequency component of the electric field. The polarization and electric field vectors are given by
\begin{align}
& \mathbf{P}^\pm\left(z ,\tau, v\right) = P^\pm\left(z ,\tau, v\right) e^{\pm i\omega_0\tau}\boldsymbol{\epsilon}_\mathrm{d}\label{eq:penvelope}\\
& \mathbf{E^\pm}\left(z ,\tau\right) = E^\pm\left(z ,\tau\right) e^{\mp i\omega_0\tau}\boldsymbol{\epsilon}_\mathrm{d}\label{eq:Eenvelope}
\end{align}
with $\boldsymbol{\epsilon}_\mathrm{d} = \mathbf{d}/d$ the unit polarisation vector associated with the molecular transition at a frequency $\omega_0 = c k$ and a transition dipole moment $d=\left| \mathbf{d}\right|$. The phenomenological time-scales $T_1$ and $T_2$ respectively account for (non-coherent) relaxation of the population inversion and de-phasing of the polarisation of the system. The temporal evolution of our systems is initiated through internal fluctuations in $n^\prime$ and $P^+$ modelled with an initial non-zero Bloch angle (see \citealt{Rajabi2019} for more details). The polarisation pump $\Lambda_P$ in equation (\ref{eq:dP/dt}) consists entirely of those fluctuations. The same results would be obtained if a constant background seed signal was instead applied at the input of the systems, although this would reduce the length $L$ by a small fraction (i.e., $\sim10\%$ to $20\%$). 

We note that $\Lambda_n$ in equation (\ref{eq:dN/dt}) is not meant to represent the physical pump of the system. It is rather an ``effective pump'' to phenomenologically account for the different group of molecular transitions and interactions with the surroundings that lead to a net change in the population density of the two-level system. In our analysis, the system instantaneously `feels' the effect of this effective pump but its response to it takes place over the time-scale $T_1$ at low field intensities. The same is true for the response of the polarization $P^+$ in equation (\ref{eq:dP/dt}), but with a time-scale of $T_2$ (still at low intensities).

In a similar manner, the phenomenological time-scales $T_1$ and $T_2$ are also ``effective time-scales'' adapted to the two-level molecule and MBE framework.  That is, $T_1$ and $T_2$ are respectively related to non-coherent relaxation and de-phasing between the two maser/superradiance energy levels.

\begin{figure*}
    \centering
    \begin{minipage}[t]{.48\textwidth}
    \includegraphics[width=\columnwidth]{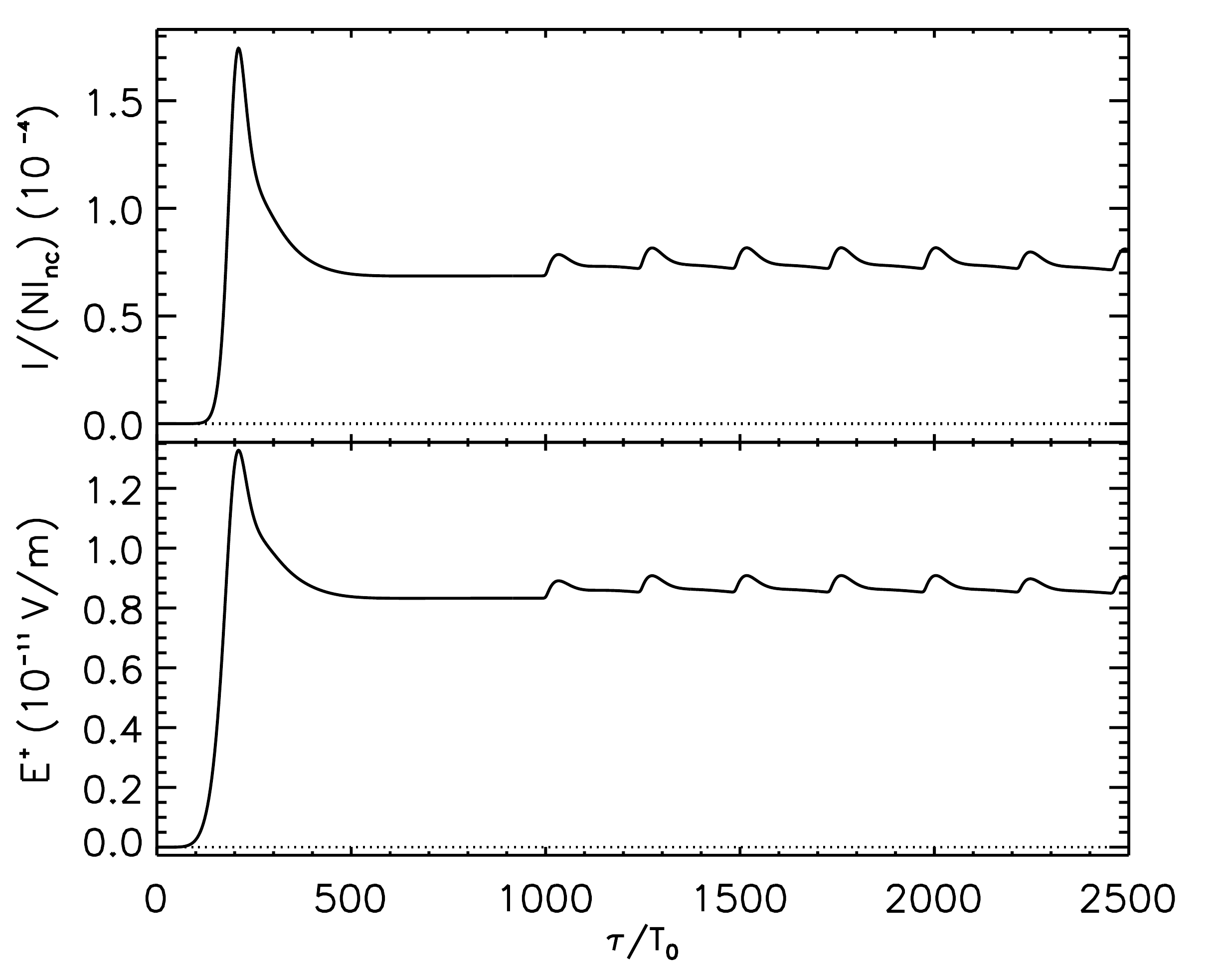}
    \end{minipage}\qquad
    \begin{minipage}[t]{.48\textwidth}
    \includegraphics[width=\columnwidth]{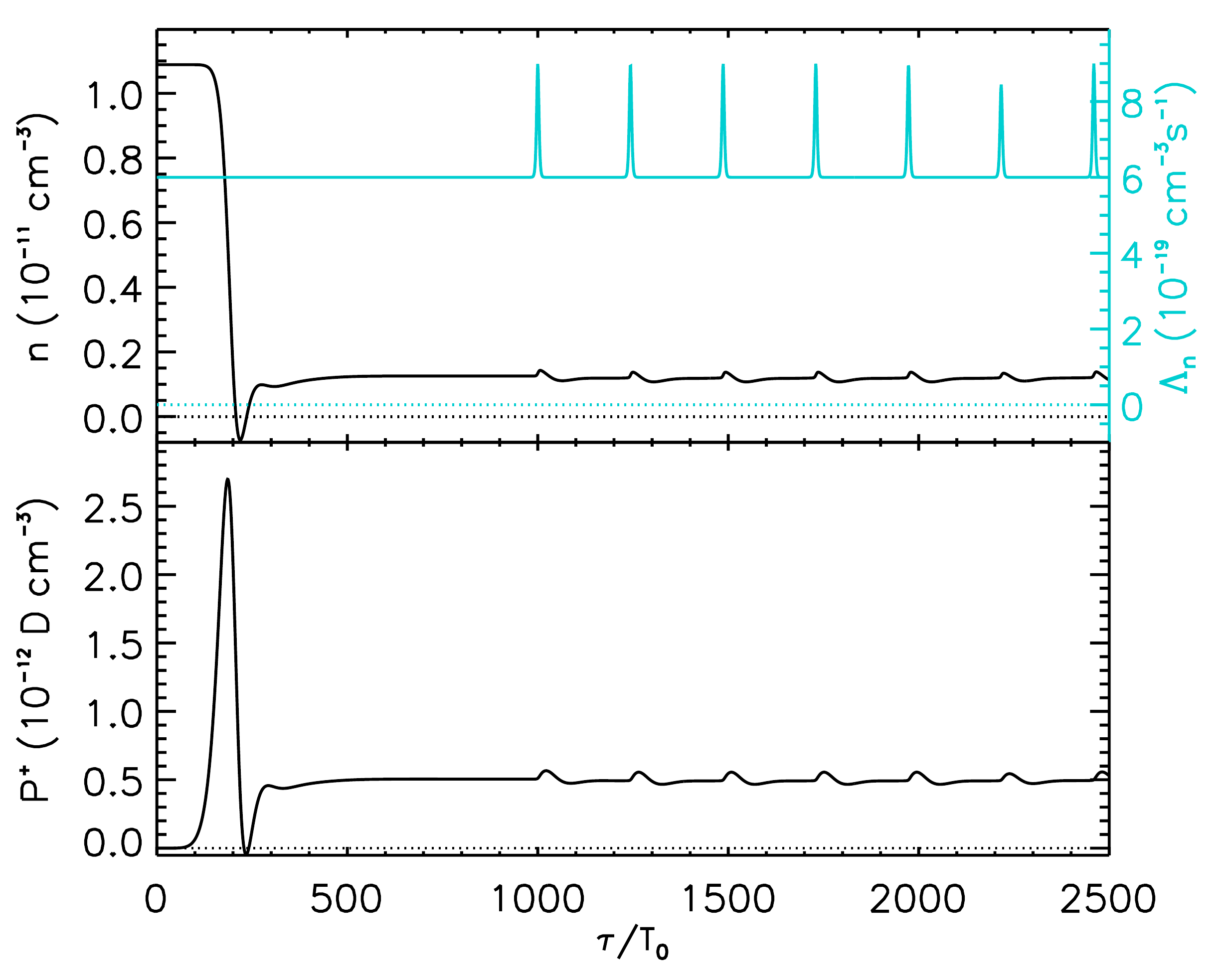}
    \end{minipage}
    \caption{Numerical solution of the Maxwell-Bloch equations (i.e., equation \ref{eq:dN/dt}-\ref{eq:dE/dz}) for the OH~1665~MHz flare of Figure \ref{fig:OH1665fit}. Temporal evolution of the intensity (left top) and the electric field (left bottom) at the end-fire ($z=L$) of a single sample in response to the initial population inversion level $n_0=1.09\times 10^{-11}$~cm$^{-3}$ (right top; black curve) and the inversion pump (right top; cyan curve). The initial population inversion, which corresponds to a level of approximately $10^{-2}$~cm$^{-3}$ for a molecular population spanning a velocity range of $\sim 1$~km~s$^{-1}$, leads to a superradiance burst after some time delay before settling in the saturated maser regime. While a constant inversion pump level $\Lambda_0=n_0/T_1$ is applied throughout, the pump pulses of amplitude $\Lambda_{\mathrm{1},m}$ are then periodically applied for $\tau\geq 1000\,T_0$ ($T_0=1$~d). Also shown is the time evolution of the polarisation (right bottom).}
    \label{fig:pulse_sequence}
\end{figure*}

Although it is possible that the observations modelled in this paper suffer from confusion in view of the limited spatial resolution with which they were acquired and that the resulting flux densities may include contributions from more than one source, we treat the data as they are and assume that the signal originates from only one source. For example, we do not separate and model independently the more-or-less constant flux density levels measured between flares and the flares themselves; all our models stem from a unique solution of the MBE. We also chose a value for the time-scale $T_\mathrm{p}$ such that the pump pulse's temporal width is significantly shorter than those of the flares themselves in order to better show the transient nature of the system's response. As mentioned earlier, just as for its shape, the chosen pump pulse's width is to a large extent arbitrary and common to all models, irrespective of the transition. In a similar manner, the length of the system $L$ is kept the same for all models.

We note that the sample's length (chosen at $L=50$~au) is largely arbitrary since, from equation (\ref{eq:TR}), it is the column density of the population inversion $nL$ that sets the superradiance time-scale. Any variation in $L$ can be absorbed through a corresponding change in $n$. In a similar manner, our analyses are applicable to any portion of a spectral line since in real systems the so-called Arecchi-Courtens condition \citep{Arecchi1970} serves to auto-regulate the superradiance time-scale across a spectral line through coupling between velocity slices within the population inversion (see \citealt{Wyenberg2022} for a detailed discussion).  

The values for the constant pump level $\Lambda_{\mathrm{0}}$ we converged to while fitting the data all lead to quasi-steady states (i.e., in between flares) being, to varying degrees, in the saturated maser regime. Our numerical computations are thus started with the pump level set to $\Lambda_{\mathrm{0}}$ and corresponding population inversion level $n_0=\Lambda_{\mathrm{0}}T_1$, which after a time delay leads to a superradiance burst before settling to the saturated maser regime \citep{Rajabi2020b}. The pump pulses of amplitude $\Lambda_{\mathrm{1},m}$ are then periodically applied and the system's response fitted to the data for at least two of these pulses; an example is shown in Figure \ref{fig:pulse_sequence} for the OH~1665~MHz fit, at $+1.7 =$~km~s$^{-1}$, of Figure \ref{fig:OH1665fit}. The model fits are scaled to the data.

Finally, and as explained in \citet{Rajabi2019} (see their Sec. 3.1), our numerical calculations pertain to systems that are of small cross-section (in an astronomical context) in view of the coherent nature of superradiance systems and could not be spatially resolved through observations. We call these entities samples. As will be discussed in Sec. \ref{sec:discussion}, a physical flaring maser-hosting region in G9.62+0.20E of typical size (e.g., 1--10~au) would therefore host a large number of samples that would more or less simultaneously erupt and together yield the measured flux densities. 

\subsection{OH~1665~MHz}\label{sec:1665MHz}

\begin{figure}
    \centering
    \includegraphics[width=0.47\textwidth]{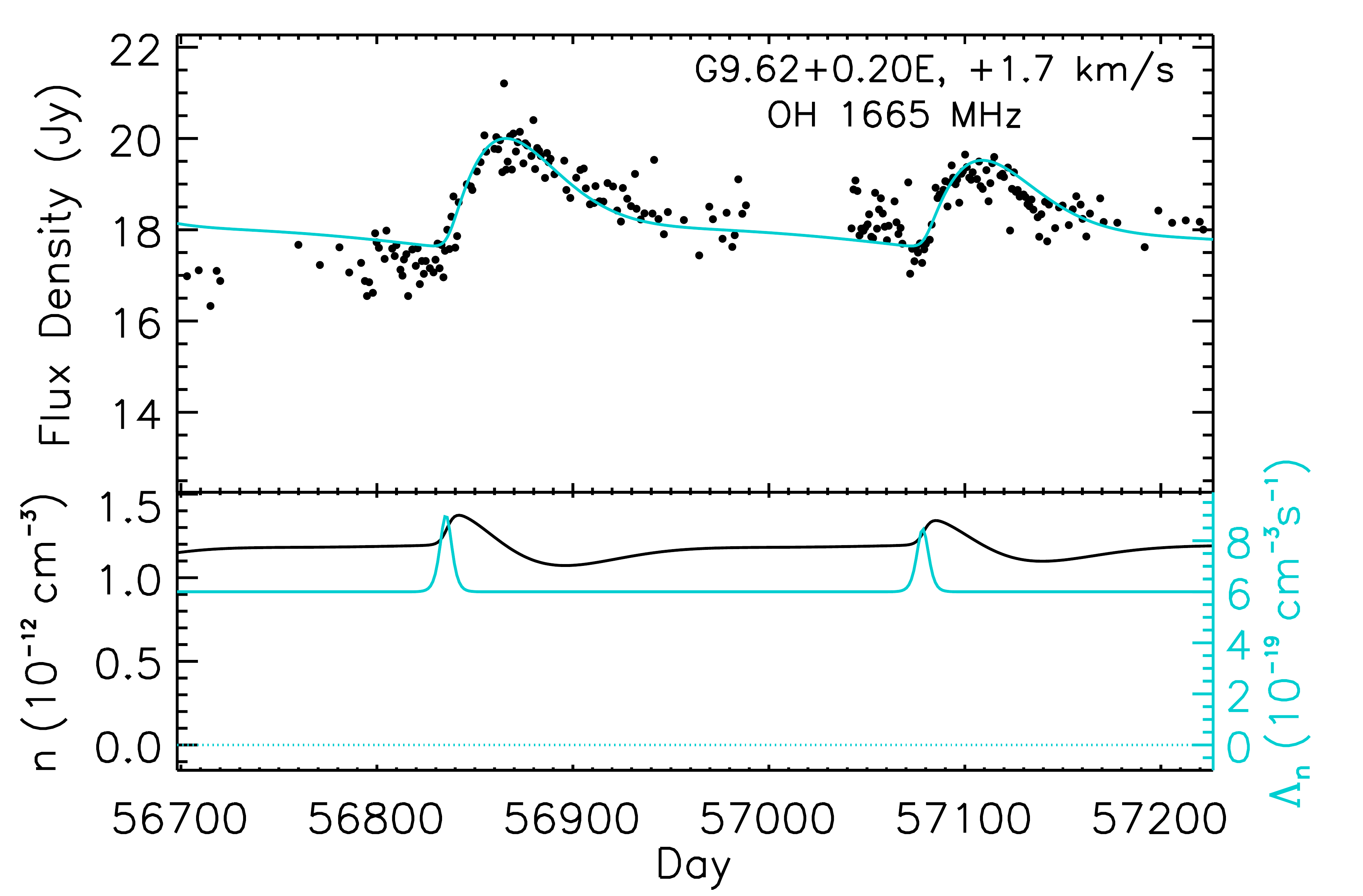}
    \caption{Top: Model fit (light/cyan curve) to the OH~1665~MHz data at $v_{\mathrm{lsr}} = +1.7$~km~s$^{-1}$ (black dots). Bottom: The population inversion (solid black curve; left vertical scale) is initially set to $n_0 = 1.09 \times 10 ^{-11}$~cm$^{-3}$ (see Fig. \ref{fig:pulse_sequence}), which corresponds to a level of approximately $10^{-2}$~cm$^{-3}$ for a molecular population spanning a velocity range of $\sim 1$~km~s$^{-1}$. The pump signal $\Lambda_n$ (cyan curve; right vertical scale) consists of the constant pump rate $\Lambda_{\mathrm{0}} = 6\times 10^{-19}$~cm$^{-3}$s$^{-1}$ and a series of longitudinal pump pulses of amplitude $\Lambda_{\mathrm{1},m}$, which are responsible for the presence of the two transient flares ($\Lambda_{\mathrm{1},m} = 3\times 10^{-19}$~cm$^{-3}$s$^{-1}$ for the first flare). The length for this system, and all others presented in this paper, is set to $L=7.5\times 10^{14}$~cm (i.e., 50~au), while pump pulse duration is $T_{\mathrm{p}} = 3.5 \times 10^{5}$~s. The relaxation, dephasing and superradiance time-scales are $T_1 = 1.8 \times 10^7$~s, $T_2 = 1.2 \times 10^6$~s and $T_{{\mathrm{R}}_0} = 4.5 \times 10^{4}$~s, respectively.} 
    \label{fig:OH1665fit}
\end{figure}

In Figure \ref{fig:OH1665fit}, we show the results of our model for the OH~1665~MHz line. In the top panel, the light curve for the velocity channel at $v_{\mathrm{lsr}} = +1.7$~km~s$^{-1}$ is shown, where the black dots represent the data. The population inversion is initially set to $n_0 = 1.09 \times 10 ^{-11}$~cm$^{-3}$, which corresponds to a level of approximately $10^{-2}$~cm$^{-3}$ for a molecular population spanning a velocity range of $\sim 1$~km~s$^{-1}$. The model fit (cyan curve) is produced by solving equations (\ref{eq:dN/dt})-(\ref{eq:dE/dz}) for a sample of length $L=7.5\times 10^{14}$~cm (i.e., 50~au) while applying a pump signal in the form of equation (\ref{eq:pump}) propagating longitudinally along the symmetry axis. In the bottom panel, we show the temporal evolution of the population inversion $n$ (black curve; $n^\prime=n/2$ in equations \ref{eq:dN/dt}-\ref{eq:dE/dz}) at the end-fire $z = L$ using the vertical scale on the left and the total pump signal $\Lambda_n$ (cyan curve) using the vertical scale on the right. The total pump signal consists of the constant pump rate $\Lambda_{\mathrm{0}} = 6\times 10^{-19}$~cm$^{-3}$s$^{-1}$ and a series of longitudinal pump pulses of amplitude $\Lambda_{\mathrm{1},m}$. The constant level $\Lambda_{\mathrm{0}}$ accounts for the presence of the quasi-steady state maser regimes between the transients and we converged to a pump pulse value $\Lambda_{\mathrm{1},m} = 3\times 10^{-19}$~cm$^{-3}$s$^{-1}$ to fit the first flare on the top panel, while the second required a pulse $0.82$ times that level. The pump pulse duration is set to $T_{\mathrm{p}} = 3.5 \times 10^{4}$~s (or 4~d), as for all other cases to follow. Since $n_0 = \Lambda_0 T_1$ we find $T_1 = 1.8 \times 10^7$~s from the aforementioned values of $n_0$ and $\Lambda_0$.

As discussed earlier, after the superradiance is initially triggered the system settles into a quasi-steady state (see Figure \ref{fig:pulse_sequence} at $500\,T_0\lesssim\tau<1000\,T_0$) while retaining a relatively high degree of polarization. The later arrival of excitation pump pulses triggers further coherent bursts of radiation. With periodic pump pulses this scenario can repeat, where each event is followed by a quasi-steady state maser regime. However, this can only take place if the system is not too strongly disturbed by dephasing processes (e.g., elastic collisions) as those can actively weaken the polarization and lower the coherence levels in the system below those required for superradiance. Here, we find $T_2 = 1.2 \times 10^6$~s which is almost a factor of 15 shorter than $T_1$. In earlier papers \citep{Rajabi2016A,Rajabi2016B,Rajabi2019, Rajabi2020} where non-periodic flaring events were discussed, it was emphasized that the characteristic time-scale of superradiance $T_{\mathrm{R}}$ must be shorter than both $T_1$ and $T_2$. While $T_{\mathrm{R}}$ is given by equation (\ref{eq:TR}) for an isolated burst event, it is not a well-defined parameter for the case of periodic and repeating superradiance bursts. We can, however, focus on the superradiance burst that takes place before the periodic flaring phase in our numerical calculations and define $T_{{\mathrm{R}}_0}$ using the initial inverted column density $\left(n_0L\right)$ right before the onset of the initial superradiance burst. Here, we have $\left(n_0L\right) = 8.16 \times 10 ^{3}$~cm$^{-2}$ and $T_{{\mathrm{R}}_0} = 4.5 \times 10^{4}$~s, which is shorter than $T_1$ and $T_2$ as expected.

An important feature to note in Figure \ref{fig:OH1665fit} is the evident mismatch between the pump pulses duration and bursts duration. This points out to a transient phase in the response of the system. When a fast pump pulse of short duration $T_{\mathrm{p}}$ interacts with the system, it forces a rapid increase in the population inversion level, as can be observed in the bottom panel of Figure \ref{fig:OH1665fit}. As the population inversion level reaches its peak value it triggers a transient superradiance response in the system. The duration of the superradiance burst depends on a number of parameters, among which are the physical conditions of the radiating system (i.e., the different time-scales $T_1$, $T_2$ and $T_\mathrm{R}$) and the pump pulse's amplitude. In contrast to the case of a saturated maser that mimics slow changes in the inversion pump superradiance will manifest itself in a variety of time-scales as a function of the aforementioned parameters. As seen in the top panel of Figure \ref{fig:OH1665fit} the bursts last longer than their corresponding pump pulses, which is a feature of superradiance.

\subsection{Methanol~6.7~GHz}\label{sec:6.7GHz}

The periodic flaring of the methanol~6.7~GHz line in G9.62+0.20E is intriguing for a few reasons. As stated in Sec. \ref{sec:Introduction}, it has already been established by \citet{MacLeod2022} that a second period of 52~d is also observed at 8.8~km~s$^{-1}$ in this transition, while we report here and later in Sec. \ref{sec:unusual} on further differences observed at other velocities. Accordingly, we model the light curves for two different velocity channels, i.e., $v_{\mathrm{lsr}} = +5$~km s$^{-1}$ and $+8$~km s$^{-1}$, in this transition here. 

\begin{figure}
    \centering
    \includegraphics[width=0.47\textwidth]{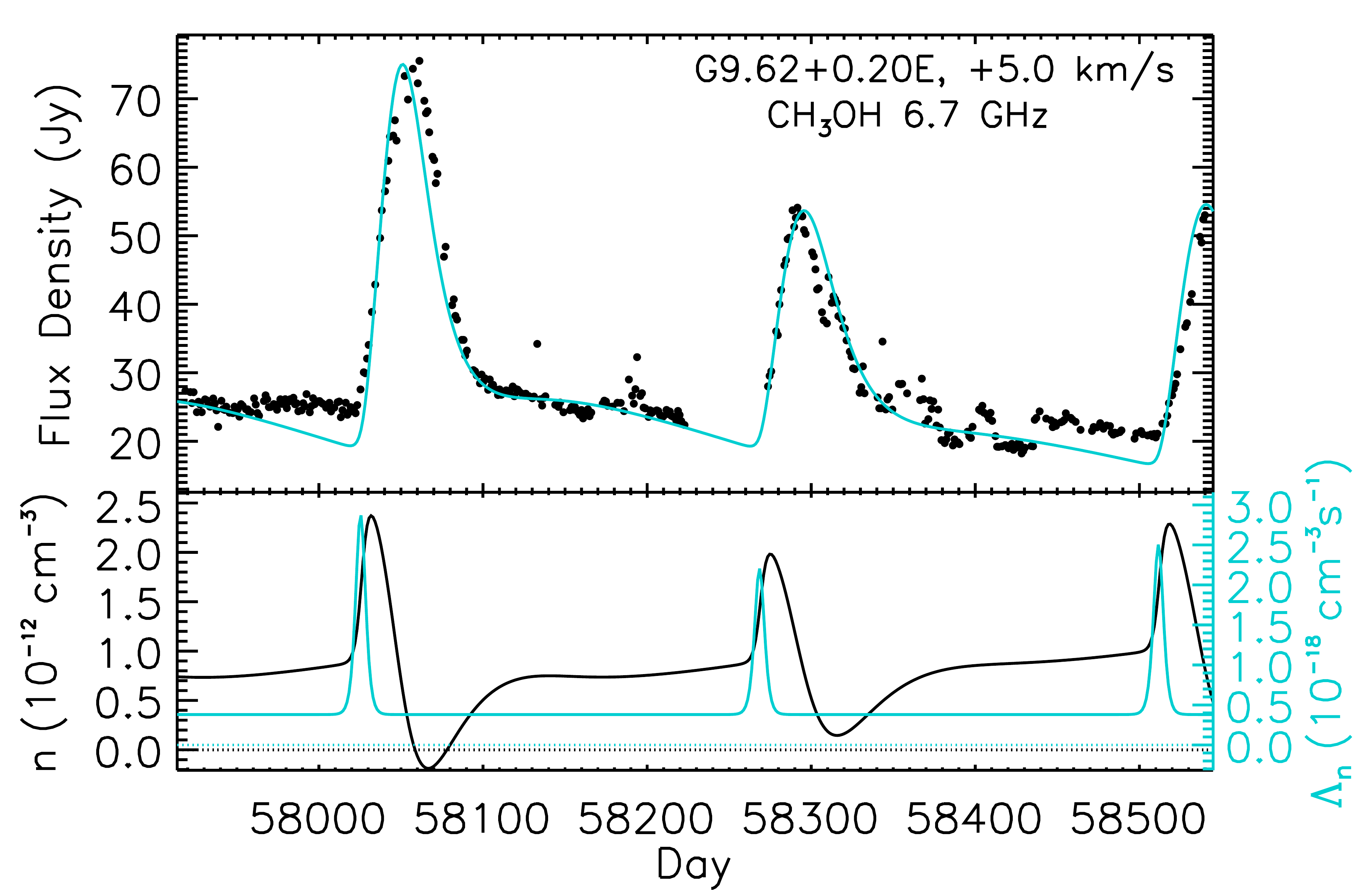}
    \caption{Same as Figure \ref{fig:OH1665fit} but for the methanol~6.7~GHz feature at $v_{\mathrm{lsr}} = +5$~km s$^{-1}$. For this system we set $\Lambda_{\mathrm{0}} = 3.8 \times 10^{-19}$~cm$^{-3}$s$^{-1}$, $\Lambda_{1,\mathrm{m}} = 2.5\times 10^{-18}$~cm$^{-3}$s$^{-1}$ for the first flare, $T_1 = 1.9 \times 10^7$~s, $T_2 = 1.1 \times 10^6$~s and $T_{{\mathrm{R}}_0} = 4.9 \times 10^{4}$~s.}
    \label{fig:Methanol6.7-5kmfit}
\end{figure}

In Figure \ref{fig:Methanol6.7-5kmfit} we show the data for $v_{\mathrm{lsr}} = +5$~km s$^{-1}$ in the top panel along with the solid cyan curve for the fit generated using our model discussed in Sec. \ref{sec:SRvsMaser}. As before, the bottom panel of the figure details the time evolution of the population inversion $n$ (black curve) using the left vertical scale and, on the right, the total pump signal (cyan curve). 

Once again, the pump signal is assumed to act longitudinally propagating along the symmetry axis of the system and has the same duration $T_{\mathrm{p}}$ (see equation \ref{eq:pump}) as that used for the OH~1665~MHz line analysis. In order to fit the data for the flares shown in the top panel of Figure \ref{fig:Methanol6.7-5kmfit}, the constant pump rate is set to $\Lambda_{\mathrm{0}} = 3.8 \times 10^{-19}$~cm$^{-3}$s$^{-1}$ and the amplitude of the first pump pulse is set to $\Lambda_{1,\mathrm{m}} = 2.5\times 10^{-18}$~cm$^{-3}$s$^{-1}$, with that of the second and third pulses to 0.73 and 0.85 times this level, respectively. Note that the required pump pulse amplitudes are almost an order of magnitude larger than those used for the OH~1665~MHz analysis. This is expected considering the higher amplitude of the methanol~6.7~GHz flares relative to the quasi-steady state level compared to those seen in Figure \ref{fig:OH1665fit} for the OH~1665~MHz line. The arrival of each narrow pump pulse once again results in a sharp increase in the population inversion $n$. This triggers a transient behaviour in the system where energy is released efficiently, collapsing the population inversion to negative values, a characterizing feature of superradiance. After this phase, the constant pump level restores $n$ to the positive pre-pump pulse levels where the system evolves in the quasi-steady state saturated maser regime. 

Our analysis results in an initial column density $\left(n_0L\right) = 5.4 \times 10 ^{3}$~cm$^{-2}$ corresponding to a superradiance time-scale $T_{{\mathrm{R}}_0} = 4.9 \times 10^{4}$~s. We also find $T_1 = 1.9 \times 10^7$~s and $T_2 = 1.1 \times 10^6$~s, which are very similar to those found in the previous section for the OH~1665~MHz line example.
\begin{figure}
    \centering
    \includegraphics[width=0.47\textwidth]{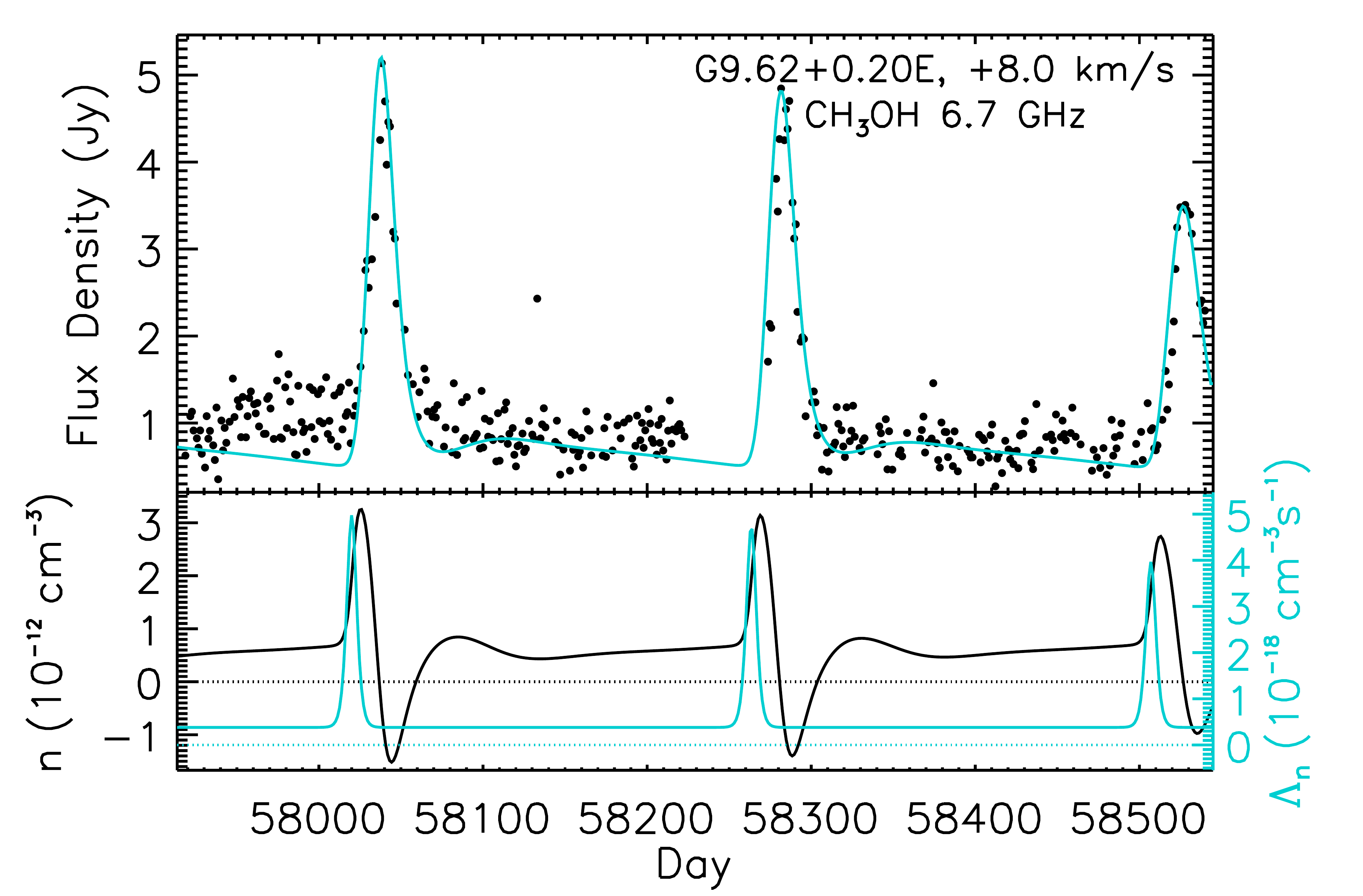}
    \caption{Same as Figure \ref{fig:Methanol6.7-5kmfit} but for the methanol~6.7~GHz feature at $v_{\mathrm{lsr}} = +8$~km s$^{-1}$. The parameters for this system are the same as for the $v_{\mathrm{lsr}} = +5$~km s$^{-1}$ feature in Figure \ref{fig:Methanol6.7-5kmfit}, except for an increase for the pump pulses with $\Lambda_{1,\mathrm{m}} = 4.6\times 10^{-18}$~cm$^{-3}$s$^{-1}$ for the first flare. This leads to a correspondingly shorter superradiance time-scale $T_{{\mathrm{R}}_0}$ that accounts for \textit{both} the flares' shorter duration and their more pronounced amplitudes relative to the quasi-steady state flux levels in between flares.}
    \label{fig:Methanol6.7-8kmfit}
\end{figure}

In Figure \ref{fig:Methanol6.7-8kmfit}, we show one more flare in the methanol~6.7~GHz line, this time for the velocity channel $v_{\mathrm{lsr}} = +8$~km s$^{-1}$. The flares at this velocity channel are not as strong as those for $v_{\mathrm{lsr}} = +5$~km s$^{-1}$ with flux density peaks around 5~Jy compared to approximately 70~Jy for those seen for $v_{\mathrm{lsr}} =+5$~km s$^{-1}$ in Figure \ref{fig:Methanol6.7-5kmfit}. The quasi-steady state flux levels are also lower ($\approx 1$~Jy compared to $\approx 25$~Jy for $v_{\mathrm{lsr}} = +5$~km s$^{-1}$). But the most interesting feature for our study is the duration of these flares. More precisely, the flares at $v_{\mathrm{lsr}} =+8$~km s$^{-1}$ are almost half as long as those at $v_{\mathrm{lsr}} =+5$~km s$^{-1}$. For saturated masers one would expect that flares from the same molecular species and transition reacting to changes in a common pump would exhibit similar durations. This is contrary to what we observe here. 

 This behaviour, however, is consistent with and easily accounted for within the framework of superradiance. That is, the flares shown in Figure \ref{fig:Methanol6.7-8kmfit} are fitted by only changing the pump pulse amplitudes while leaving the other parameters the same as those for $v_{\mathrm{lsr}} = +5$~km s$^{-1}$. More precisely, the amplitude of the first two pump pulses is set to $\Lambda_{1,\mathrm{m}} = 4.6\times 10^{-18}$~cm$^{-3}$s$^{-1}$ and that of the third to 0.82 times this level. Furthermore, and as seen in the figure, adjusting the pump pulse amplitudes to these higher values not only reproduces the narrower flare duration seen for $v_{\mathrm{lsr}} = +8$~km s$^{-1}$ but also captures the different burst peak amplitudes relative to the quiescent levels measured between them. In Figure \ref{fig:Methanol6.7-5kmfit}, the relative burst peak amplitude to quasi-steady state maser level is about a factor of 3, while for $v_{\mathrm{lsr}} = +8$~km s$^{-1}$ there is a 5-fold increase. By nearly doubling the pump pulse amplitude from $v_{\mathrm{lsr}} = +5$~km s$^{-1}$ to $v_{\mathrm{lsr}} = +8$~km s$^{-1}$ to fit for the narrower bursts we also accounted for their relative increase in amplitude. We also note that a similar result is obtained by simply increasing the length of the sample by about $30\%$ while keeping all other parameters unchanged (including the pump). The increase in the column density of the inverted population then leads to a shorter superradiance time-scale that matches the $v_{\mathrm{lsr}} = +8$~km s$^{-1}$ data well.  

\subsection{Methanol~12.2~GHz}\label{sec:12.2GHz}

\begin{figure}
    \centering
    \includegraphics[width=0.47\textwidth]{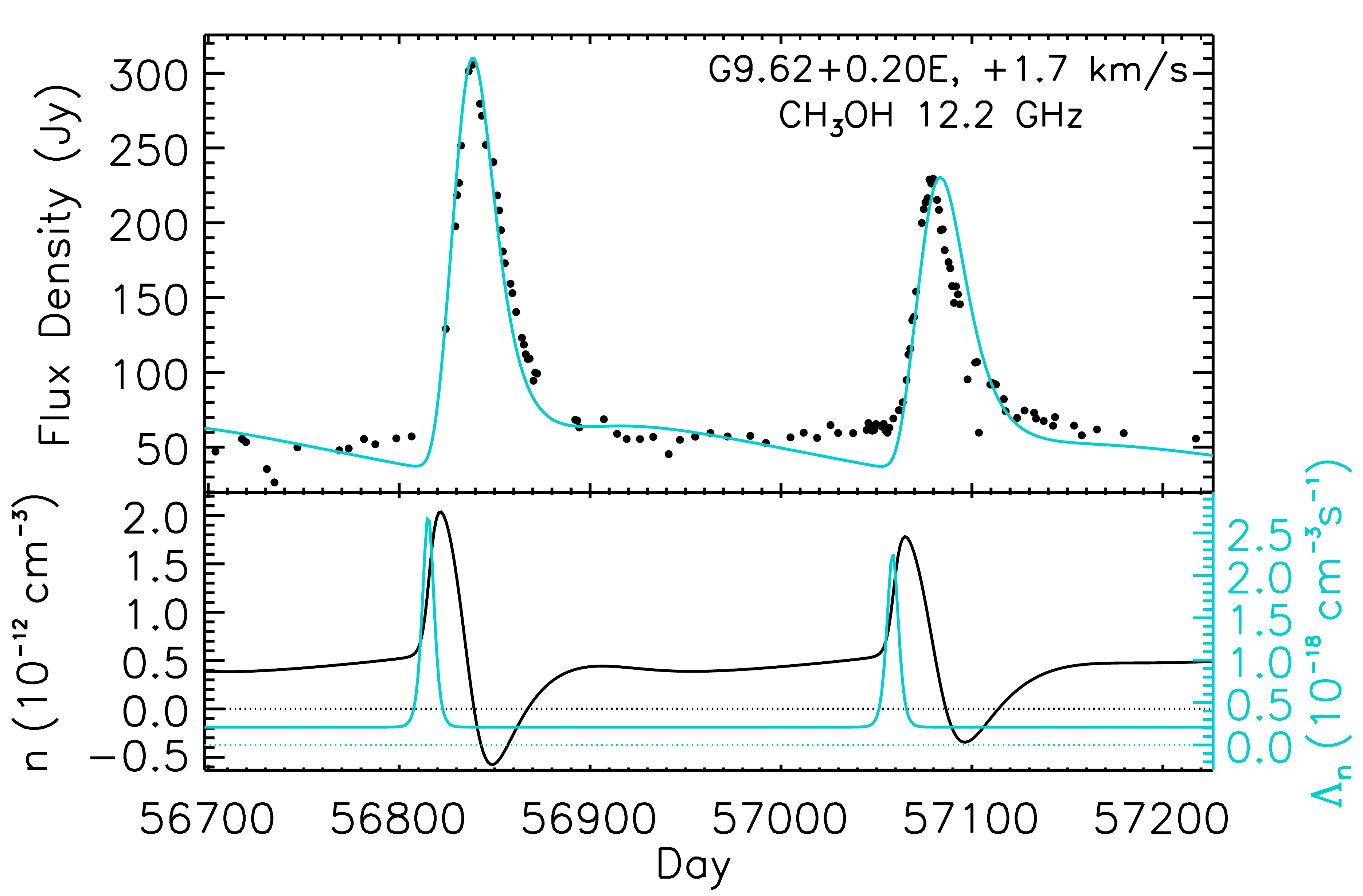}
    \caption{Same as Figure \ref{fig:OH1665fit} but for the methanol~12.2~GHz feature at $v_{\mathrm{lsr}} = +1.7$~km s$^{-1}$. For this system we set $\Lambda_{\mathrm{0}} = 2.1 \times 10^{-19}$~cm$^{-3}$s$^{-1}$, $\Lambda_{1,\mathrm{m}} = 2.5\times 10^{-18}$~cm$^{-3}$s$^{-1}$ for the first flare, $T_1 = 1.9 \times 10^7$~s, $T_2 = 1.1 \times 10^6$~s and $T_{{\mathrm{R}}_0} = 3.0 \times 10^{4}$~s.}
    \label{fig:Methanol12fit}
\end{figure}

In Figure \ref{fig:Methanol12fit}, we show the light curve of the methanol~12.2~GHz at $v_{\mathrm{lsr}} = +1.7$~km s$^{-1}$ as an example. Similar to Figures \ref{fig:OH1665fit}, \ref{fig:Methanol6.7-5kmfit} and \ref{fig:Methanol6.7-8kmfit}, in the top panel, we show fits (cyan curve) from our model for the data (black dots), while in the bottom panel we show the total pump signal (cyan curve) used in our analysis along with the time evolution of the population inversion $n$ (black curve) for sample of length $L=7.5\times 10^{14}$~cm, similar to what we use for all our analyses.

As seen in the figure, the flares in this line are quite strong with the peak flux density of the first flare reaching 300~Jy. That is, a 6-fold increase from the quasi-steady state maser level. The pump signal used to model the data is composed of a constant pump rate $\Lambda_{\mathrm{0}} = 2.1 \times 10^{-19}$~cm$^{-3}$s$^{-1}$ and pump pulses of nominal amplitude $\Lambda_{1,\mathrm{m}} = 2.5\times 10^{-18}$~cm$^{-3}$s$^{-1}$ (but 0.82 times this level for the last), while the duration of the pump pulses $T_{\mathrm{p}}$ is identical to that used in the previous sections. Once again we could reproduce the flare profiles and capture their peak intensities and duration by only adjusting the pump levels, while keeping the other parameters the same as for the methanol~6.7~GHz line samples (i.e., $T_1 = 1.9 \times 10^{7}$~s and $T_2 = 1.1 \times 10^6$~s). The initial inverted column density $\left(n_0L\right) = 3.0 \times 10^{3}$~cm$^{-2}$ corresponds to a superradiance time-scale $T_{{\mathrm{R}}_0} = 5.5 \times 10^{4}$~s. 

\subsection{Unusual methanol 6.7~GHz and OH~1667~MHz flares}\label{sec:unusual}

\begin{figure}
    \centering
    \includegraphics[width=0.47\textwidth]{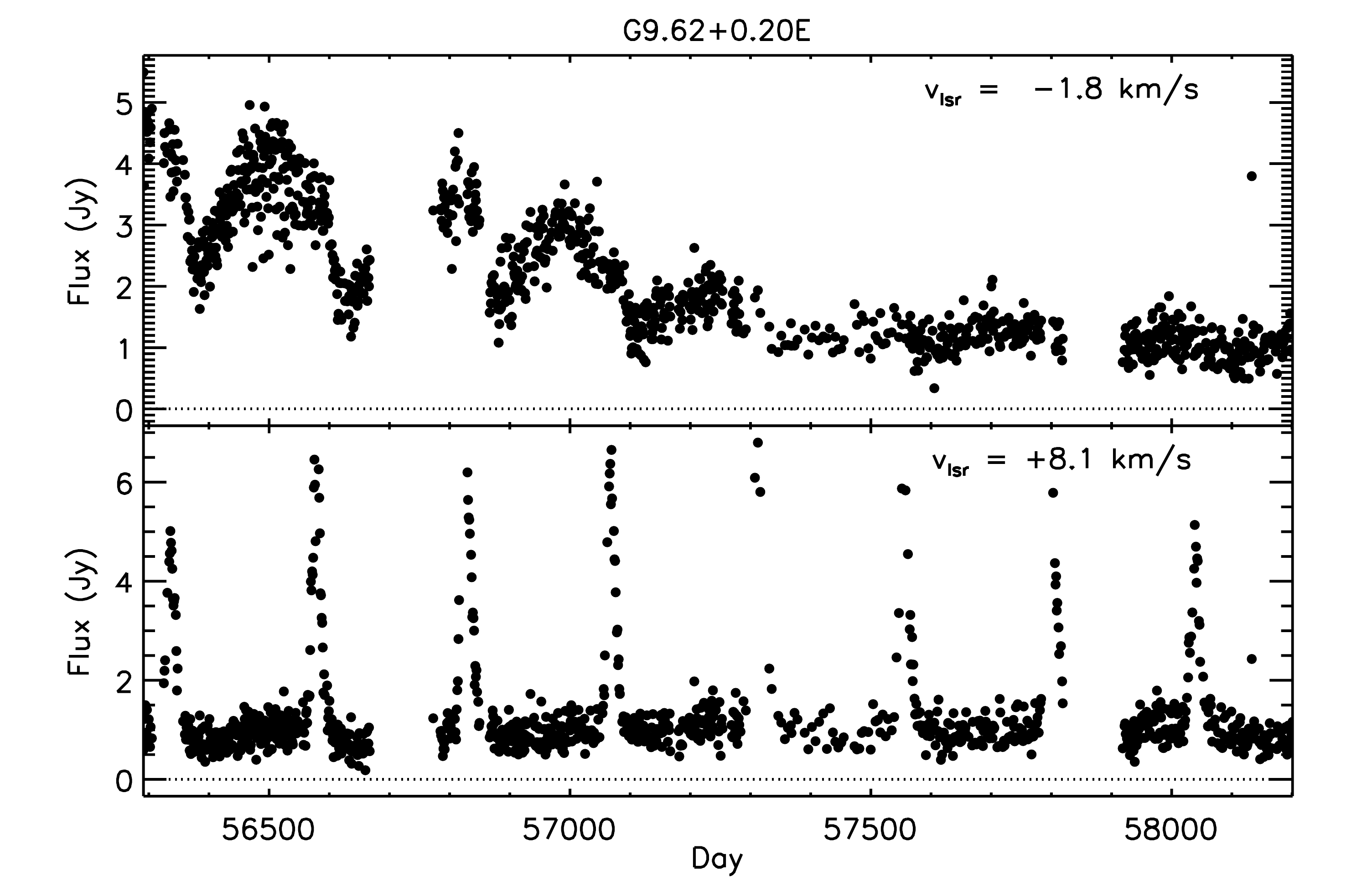}
    \caption{Time series for the methanol~6.7~GHz transition at $v_{\mathrm{lsr}} = -1.8$~km s$^{-1}$ (top) and $v_{\mathrm{lsr}} = +8.0$~km s$^{-1}$ (bottom), which was previously analyzed in Figure \ref{fig:Methanol6.7-8kmfit}. The $v_{\mathrm{lsr}} = -1.8$~km s$^{-1}$ feature is unusual in the sense that its flares seem to be completely out of sync with those at $v_{\mathrm{lsr}} = +8.0$~km s$^{-1}$ or other velocities for that line. More precisely, the flares for these two velocity channels seem to be alternating where the peak in one coincide with a dip in the other.} 
    \label{fig:methanol-1.8_8.0_series}
\end{figure}

In Figures \ref{fig:Methanol6.7-5kmfit} and \ref{fig:Methanol6.7-8kmfit}, we showed the so-called typical flares of the methanol 6.7~GHz line in G9.62+0.20E, where the flux densities peak up when the pump pulses boost the population inversion above the steady-state levels. These flares also seem to be synchronized at different velocity channels, i.e., peak up in flux density and dip out at approximately the same times. However, an unusual behaviour is observed at $v_{\mathrm{lsr}} = -1.8$~km s$^{-1}$ for which the methanol 6.7~GHz flares seem to be completely out of sync with other velocity channels. In Figure \ref{fig:methanol-1.8_8.0_series} we show the light curves for $v_{\mathrm{lsr}} = -1.8$~km s$^{-1}$ (top panel) and $v_{\mathrm{lsr}} = +8$~km~s$^{-1}$ (bottom panel) for better comparison over several periods of flaring starting at approximately MJD~56291. As seen in the figure, the flares for these two velocity channels seem to be alternating where the peak in one coincide with a dip in the other. 

Alternating behaviour has previously been observed between flares of different species. For instance, and as was mentioned in Sec. \ref{sec:Introduction}, \cite{Szymczak2016} reported periodic and alternating methanol~6.7~GHz and water~22~GHz flares in G107.298+5.639. More precisely, when water~22~GHz flares up in this source, methanol~6.7~GHz enters a quiescent phase and the dip in one is concurrent with the peak in the other. For this object, some sources for the two species are thought to be within the same gas volume of $\sim 30$--$80$~au, and \cite{Szymczak2016} associates this behaviour with changes in the dust temperature and subsequently the efficiency of the corresponding pumps. That is, the arrival of a strong infrared pulse in such a gas volume is thought to increase the population inversion of and trigger a burst of energy at the radiatively pumped methanol 6.7~GHz line but quench the collisionally pumped water~22~GHz transition. Here, we entertain a similar idea where changes in the dust temperature could control the flux density of methanol~6.7~GHz at different velocity channels. More precisely, depending on the location of the host maser cloud relative to the central radiating object the dust temperature could vary sufficiently to significantly affect the population inversion levels, in some cases (e.g., at $v_{\mathrm{lsr}} = -1.8$~km~s$^{-1}$) momentarily quenching it.   

\begin{figure}
    \centering
    \includegraphics[width=0.47\textwidth]{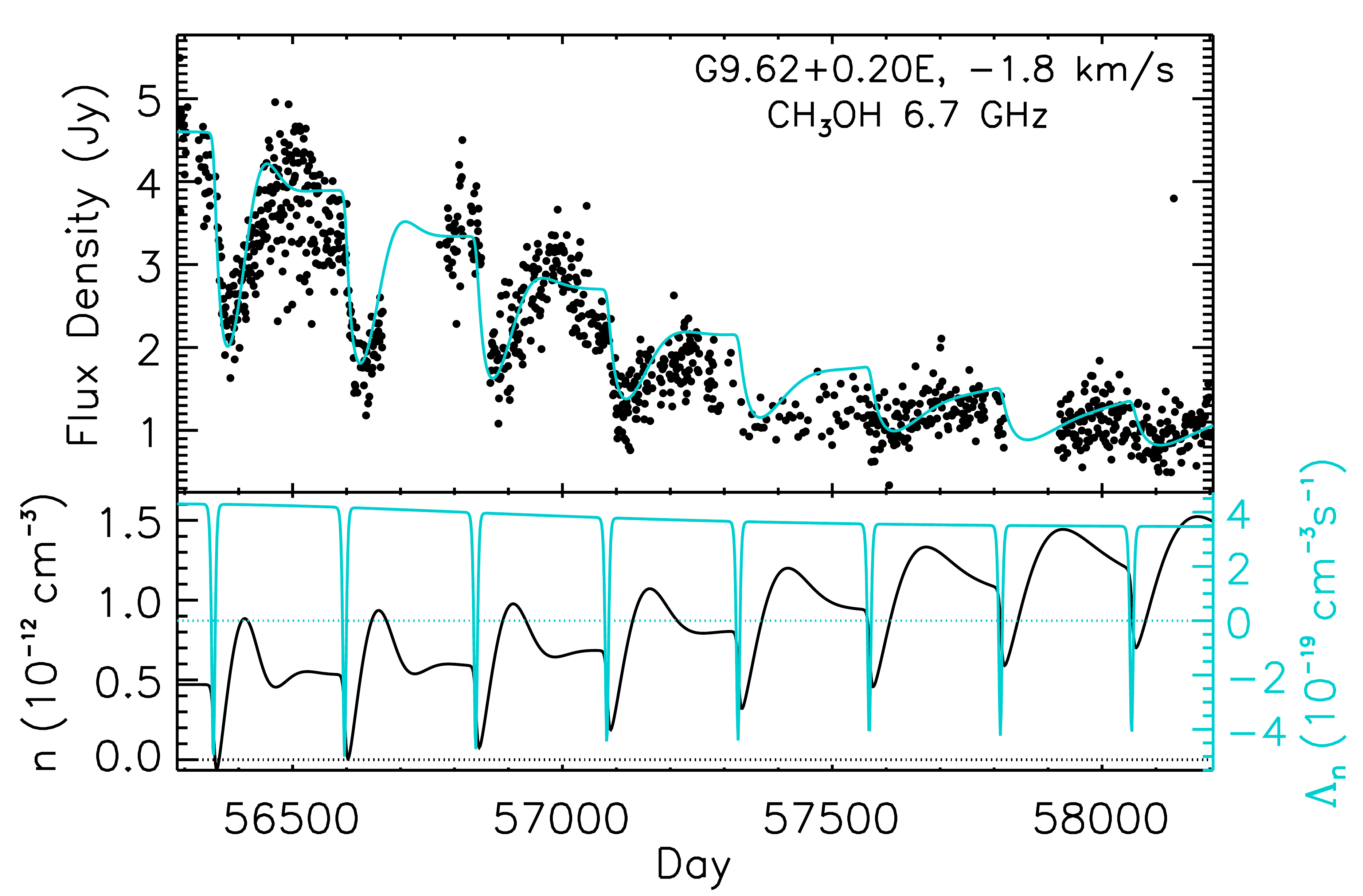}
    \caption{Same as Figure \ref{fig:OH1665fit} but for the $v_{\mathrm{lsr}} = -1.8$~km s$^{-1}$ feature in the methanol~6.7~GHz transition. For this fit the pump signal is set to the constant level $\Lambda_n\left(z,\tau\right) = \Lambda_0 = 4.3 \times 10^{-19}$~cm$^{-3}$s$^{-1}$ for $\tau< 56291$~d and to the form given in equation (\ref{eq:pump_-1.8km/s}) for $\tau\ge 56291$~d. The (negative) pump pulses have an amplitude $\Lambda_1 = -9.5 \times 10^{-19}$~cm$^{-3}$s$^{-1}$ and the slowly varying envelope brings the quasi-steady state pump level down to $0.8\Lambda_0$ over a time-scale that is 200 times longer than that of the short pump pulses. Similarly to all other fits, our model yields $T_1 = 1.9 \times 10^{7}$~s and $T_2 = 1.1 \times 10^6$~s, while we find $\left(n_0L\right) = 6.1 \times 10 ^{3}$~cm$^{-2}$ corresponding to $T_{{\mathrm{R}}_0} = 4.3 \times 10^{4}$~s.}
    \label{fig:methanol-1.8km/sfit}
\end{figure}

In Figure \ref{fig:methanol-1.8km/sfit}, we show the results of our model for methanol~6.7~GHz at $v_{\mathrm{lsr}} = -1.8$~km s$^{-1}$. In the top panel, the fit (solid curve) from our model is superposed on the data (black dots), similarly to the previous fits. In the bottom panel, the total pump signal (blue curve) is presented along with the temporal evolution of population inversion (black curve). To capture the unusual behaviour of the methanol~6.7~GHz at this velocity channel, we have set the total pump to the constant level $\Lambda_n\left(z,\tau\right) = \Lambda_0 = 4.3 \times 10^{-19}$~cm$^{-3}$s$^{-1}$ for $\tau< 56291$~d and 
\begin{align}
    \Lambda_n\left(z,\tau\right) = & \left\{\Lambda_0+\Lambda_1\sum_{m=0}^\infty \frac{1}{\cosh^{2}\left[\left(\tau-\tau_0-m\tau_1\right)/T_\mathrm{p}\right]}\right\} \nonumber \\
    & \times\left\{0.8+\frac{0.2}{\cosh^{2}\left[\tau/\left(200\, T_\mathrm{p}\right)\right]}\right\}, \label{eq:pump_-1.8km/s}
\end{align}
for $\tau\ge 56291$~d. Here, we have $\Lambda_1 = -9.5 \times 10^{-19}$~cm$^{-3}$s$^{-1}$ for the amplitude of the (negative) pump pulses and the slowly varying envelope brings the quasi-steady state pump level down to $0.8\Lambda_0$ over a time-scale that is 200 times longer than that of the short pump pulses. As is the case for all other fits the flaring period is set to $\tau_1=243.3$~d and the pump pulse duration $T_{\mathrm{p}} = 3.5\times 10^5$~s. The combination of the quasi-steady state pump and the quenching effect of the stronger short pump pulses result in a lowering of the flux of the saturated maser phases by a factor approximately 5 dropping from $\sim 5$~Jy at $\tau\simeq 56300$~d to $\sim 1$~Jy at $\tau\simeq 58000$~d.

 As seen in the bottom panel of Figure \ref{fig:methanol-1.8km/sfit}, the arrival of each negative pump pulse coincides with a dip in the population inversion which is followed by a rise in $n$ through the slow varying component of the pump, a reverse pattern to what we see in Figures \ref{fig:Methanol6.7-5kmfit} and \ref{fig:Methanol6.7-8kmfit} for other velocity channels of the methanol~6.7~GHz. As mentioned before, a rapid change in $n$ triggers a transient superradiance response in the system, which in this case is characterized by a slight overshoot pattern in $n$. This pattern can be seen in the bottom panel of Figure \ref{fig:methanol-1.8km/sfit} especially for the first few flares. 
 
 We also note that similar to all other methanol fits, our model yields $T_1 = 1.9 \times 10^{7}$~s and $T_2 = 1.1 \times 10^6$~s, while we find $\left(n_0L\right) = 6.1 \times 10 ^{3}$~cm$^{-2}$ corresponding to $T_{{\mathrm{R}}_0} = 4.3 \times 10^{4}$~s. These results obtained at $v_{\mathrm{lsr}} = -1.8$~km s$^{-1}$ are therefore consistent with our finding for other velocity channels of methanol~6.7~GHz in spite of apparent differences in the flare profiles, which can naturally be explained by the superradiance transient response to the specific profile of pump signal.   

The pattern observed for the light curve profile of the methanol~6.7~GHz at $v_{\mathrm{lsr}} = -1.8$~km s$^{-1}$ is not unique and some of the main features (a dip in the flux followed by a bump) are in common with the light curve at some velocities of the OH~1667~MHz line in G9.62+0.20E. More precisely, as initially shown in Figure \ref{fig:g9.62-all} in the second panel from the top, the flux density at $v_{\mathrm{lsr}} = +1.7$~km s$^{-1}$ of the OH~1667~MHz line also dips from the quasi-steady state levels before exhibiting a moderate flaring activity, periodically. This behaviour is not limited to $v_{\mathrm{lsr}} = +1.7$~km s$^{-1}$ shown in the figure and is also observed for a few other velocity channels of the OH~1667~MHz transition in this source (see Fig. 4 in \citealt{Goedhart2019}). As indicated in Figure \ref{fig:g9.62-all} using a vertical line at MJD 56815, the dip of the flux density of the OH~1667~MHz coincides with the start of the flaring activity in the methanol~12.2~GHz line. This is reminiscent of the alternating pattern of methanol~6.7~GHz flares at $v_{\mathrm{lsr}} = -1.8$~km s$^{-1}$ relative to those at $v_{\mathrm{lsr}} = +8$~km s$^{-1}$. As discussed before, changes in dust temperature through interaction with a strong infrared pulse radiated from a nearby object could affect the efficiency of the radiative pumps for these lines at some velocity channels depending on the location of the host clouds. As mentioned in Sec. \ref{sec:Introduction}, \citet{Goedhart2019} proposed a similar scenario based on the modulation of seed signals from free-free radiation as a function of the location from the central source to explain the dip in the OH~1667~MHz line at $v_{\mathrm{lsr}} = +1.7$~km s$^{-1}$. 

\begin{figure}
    \centering
    \includegraphics[width=0.47\textwidth]{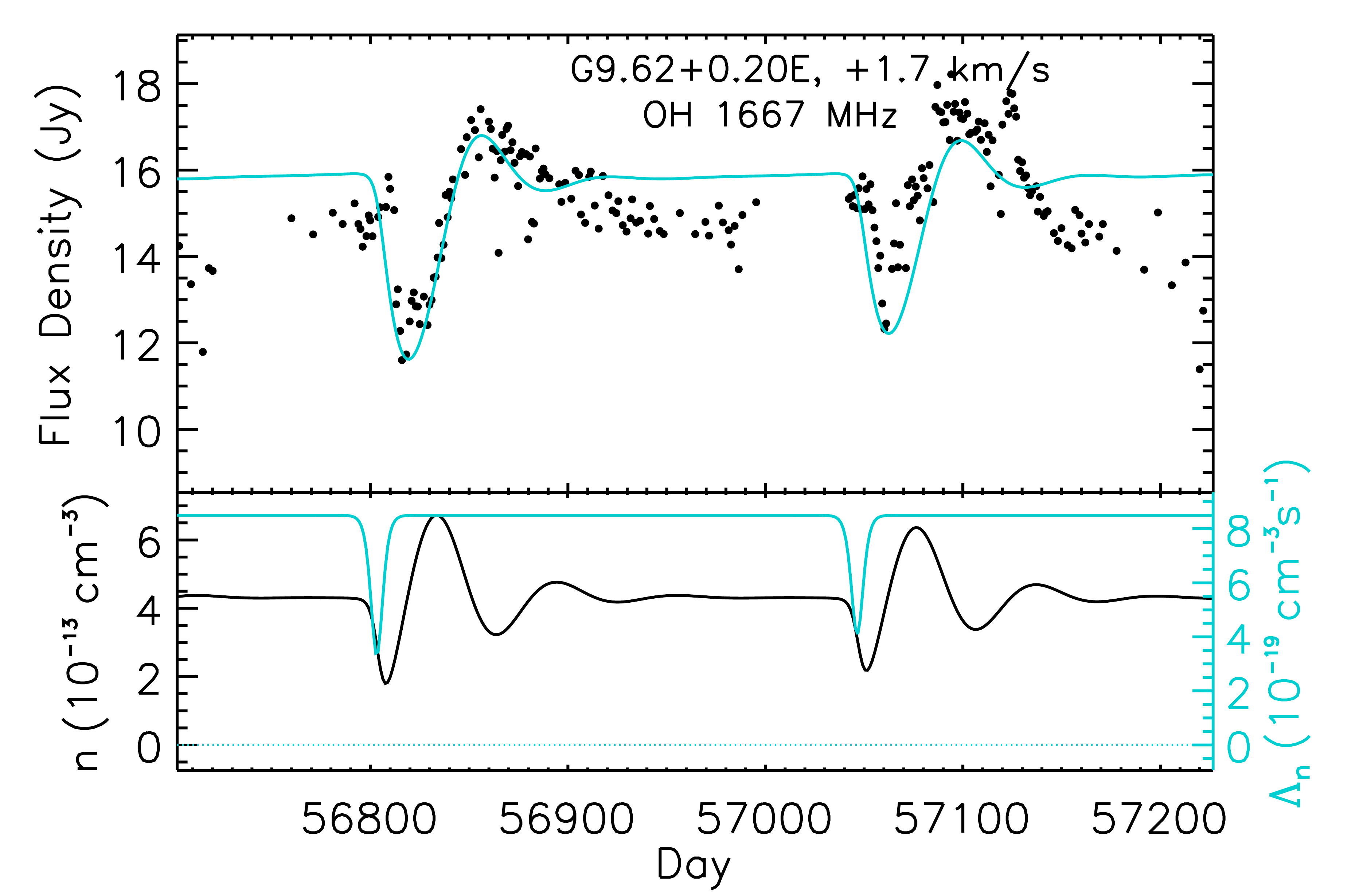}
    \caption{Same as Figure \ref{fig:OH1665fit} but for the OH~1667~GHz feature at $v_{\mathrm{lsr}} = +1.7$~km s$^{-1}$. For this system the pump signal is set to the form given in equation (\ref{eq:pump}) with $\Lambda_{\mathrm{0}} = 8.5 \times 10^{-19}$~cm$^{-3}$s$^{-1}$, $\Lambda_{1,\mathrm{m}} = -5.2\times 10^{-19}$~cm$^{-3}$s$^{-1}$ (note the negative sign) for the first flare, $T_1 = 1.9 \times 10^7$~s, $T_2 = 1.2 \times 10^6$~s and $T_{{\mathrm{R}}_0} = 2.8 \times 10^{4}$~s.}
    \label{fig:OH1667fit}
\end{figure}

As shown in the bottom panel of Figure \ref{fig:OH1667fit} (cyan curve), we implemented our pump modulation scheme by the application of pump pulses of negative amplitude superposed to a constant pump component responsible for the quasi-steady state flux levels. In the top panel of Figure \ref{fig:OH1667fit}, the light curve (black dots) along with a fit from our model (black curve) for the OH~1667~MHz line at $v_{\mathrm{lsr}} = +1.7$~km s$^{-1}$ are shown. The dip of the flux density is then seen to result from the quenching of the population inversion under the action of the (negative) pump pulses. The total pump signal shown in the figure follows equation (\ref{eq:pump}) for which $\Lambda_0 = 8.5 \times 10^{-19}$~cm$^{-3}$s$^{-1}$ and, as always, $T_{\mathrm{p}} = 3.5\times 10^5$~s. To fit the data for the first flare the amplitude of the first (negative) pump pulse is set to $\Lambda_1 = -5.2 \times 10^{-19}$~cm$^{-3}$s$^{-1}$ while that of the second is 0.85 times this level. As seen in the figure, after the population inversion level reaches its minimum, it recovers and overshoots through the re-establishing of the constant component of the pump. This also triggers a transient behaviour that manifests itself as the observed flare, after which the flux density level returns to quasi-steady state levels. Although it is certainly a possibility that the dips and flares in flux density are due to independent OH-emitting regions (i.e., one close to the central object responsible for the dips and another further away for the flares), in our model the superradiance transient response of a single system can account for both features.   

Finally, our model for the OH~1667~MHz at $v_{\mathrm{lsr}} = +1.7$~km s$^{-1}$ yields an inverted column density $\left(n_0L\right) = 1.2 \times 10 ^{4}$~cm$^{-2}$ corresponding to $T_{{\mathrm{R}}_0} = 2.8 \times 10^{4}$~s, while the time-scales $T_1$ and $T_2$ are as reported in previous sections (i.e., $T_1 = 1.9 \times 10^{7}$~s and $T_2 = 1.1 \times 10^6$~s). 

\section{Discussion}\label{sec:discussion}

As discussed before, the fits presented here are produced using the excitation of a one-dimensional sample by a periodic inversion pump composed of a constant component and a series of pulses. The values for the relaxation and dephasing time-scales vary minimally between our whole range of models for the different transitions and molecular species. More precisely, both time-scales varied by less than $\sim 7\%$ from about $T_1 = 1.9 \times 10^7$~s (220~d) and $T_2 = 1.1 \times 10^6$~s (12.6~d). As these parameters pertain to conditions external to the molecular species, the implication is that the types of environments probed by these observations in G9.62+0.20E are not too dissimilar. It is, therefore, rewarding that we were able to account for the different flare time-scales and shapes by simply adjusting the pump's steady-state and pulse levels. In other words, by fixing the pump pulse duration to $T_\mathrm{p}=3.5 \times 10^5$~s (4~d) for all cases we effectively only adjusted the degree of coupling to the pump signal. This is as would be expected under realistic conditions when a single source is responsible for the 243~d periodicity of the flares. That is, while pump related time-scales like the period and pulse duration should not change depending on which transition/species is flaring, the coupling to the pump signal can be affected by a variety of factors (e.g., geometry, relative orientation, wavelength, intervening matter, etc.). The transient nature of superradiance and its changing response as a function of the excitation signal (i.e., the inversion pump) and the transition/species at hand makes it a natural candidate to explain the diverse flaring time-scales and profiles observed. This would have been impossible to achieve by restricting our analysis to the quasi-steady state maser regime.   

More precisely, one of the main challenges for maser-only models is the observations of multiple time-scales for flares for species/transitions responding to a common pump. For instance, for the methanol~6.7~GHz line, we show flares at $v_{\mathrm{lsr}} = +8$~km s$^{-1}$ in Figure \ref{fig:Methanol6.7-8kmfit} lasting approximately 40 days, while for the same transition at $v_{\mathrm{lsr}} = +5$~km s$^{-1}$, as shown in Figure \ref{fig:Methanol6.7-5kmfit}, the flares last almost 100 days. Such significant difference in the flare duration between two velocity channels of the methanol~6.7~GHz line transition, which is known to be radiatively pumped, requires two very different pump time-scales within the maser model. More precisely, because the profile of a saturated maser closely follows that of its pump, and since the two flares are thought to be the product of the masing action, they must follow the duration of their corresponding pump excitation. That would necessitate the existence of multiple pump time-scales, mainly from infrared pulses, in G9.62+0.20E to explain each reported time-scale. The superradiance regime, on the contrary, can reproduce different flare durations using a single pump pulse time-scale.  

On the other hand, it is also important to realize that the superradiance transient response of a system provides no information on the nature of the astronomical source responsible for the pump and its periodicity. It only concerns the underlying physical process for the flares and cannot by itself discern, for example, whether the pump is the product of stellar pulsation from accretion episodes \citep{Inayoshi2013} or CWBs \citep{VanderWalt2011,VanderWalt2016}.   

As seen in Figures \ref{fig:OH1665fit}, \ref{fig:Methanol6.7-5kmfit}, \ref{fig:Methanol6.7-8kmfit}, \ref{fig:Methanol12fit}, \ref{fig:methanol-1.8km/sfit} and \ref{fig:OH1667fit}, our simple MBE one-dimensional model using a periodic inversion pump excitation successfully captures the main behaviour and characteristics of flares in G9.62+0.20E for the four transitions discussed. The fits are not perfect, however, and could be improved. This is the case, for example, in the quiescent phase in between flares where the data often show more scatter. Differences between the fits and the observed light curves can result from a range of factors. There could be multiple regions radiating simultaneously within the telescope beam, in which case the data would contain contributions from all sources. Mild flaring activity could be taking place in some regions while other areas are in the quasi-steady state regime, which would complicate the modelling of the light curve profiles during those phases. For instance, our model fits for the methanol lines (e.g., Figs. \ref{fig:Methanol6.7-5kmfit} and \ref{fig:Methanol12fit}) show slight dips right before the start of each flare that often deviate from the data trend. 

As will be further discussed later in this section, we also believe that some assumptions implicit to our model can be affecting the quality of the fits. For example, we solve the one-dimensional MBE assuming homogeneous samples, whereas more realistic systems are likely to be non-homogeneous and asymmetric. Although more complex, three-dimensional models could better capture the detailed evolution of the system. We also use an on-resonance model for our numerical calculations with the effective bandwidth is set by the duration of the flares. Including a finite bandwidth in the MBE would ``soften'' the model fits, which could in turn improve some of our results. For instance, the combination of multiple velocity channels can smooth out some of the oscillations seen in Figure \ref{fig:OH1667fit} \citep{Wyenberg2021,Wyenberg2022}. 

\begin{figure}
    \centering
    \includegraphics[width=0.47\textwidth]{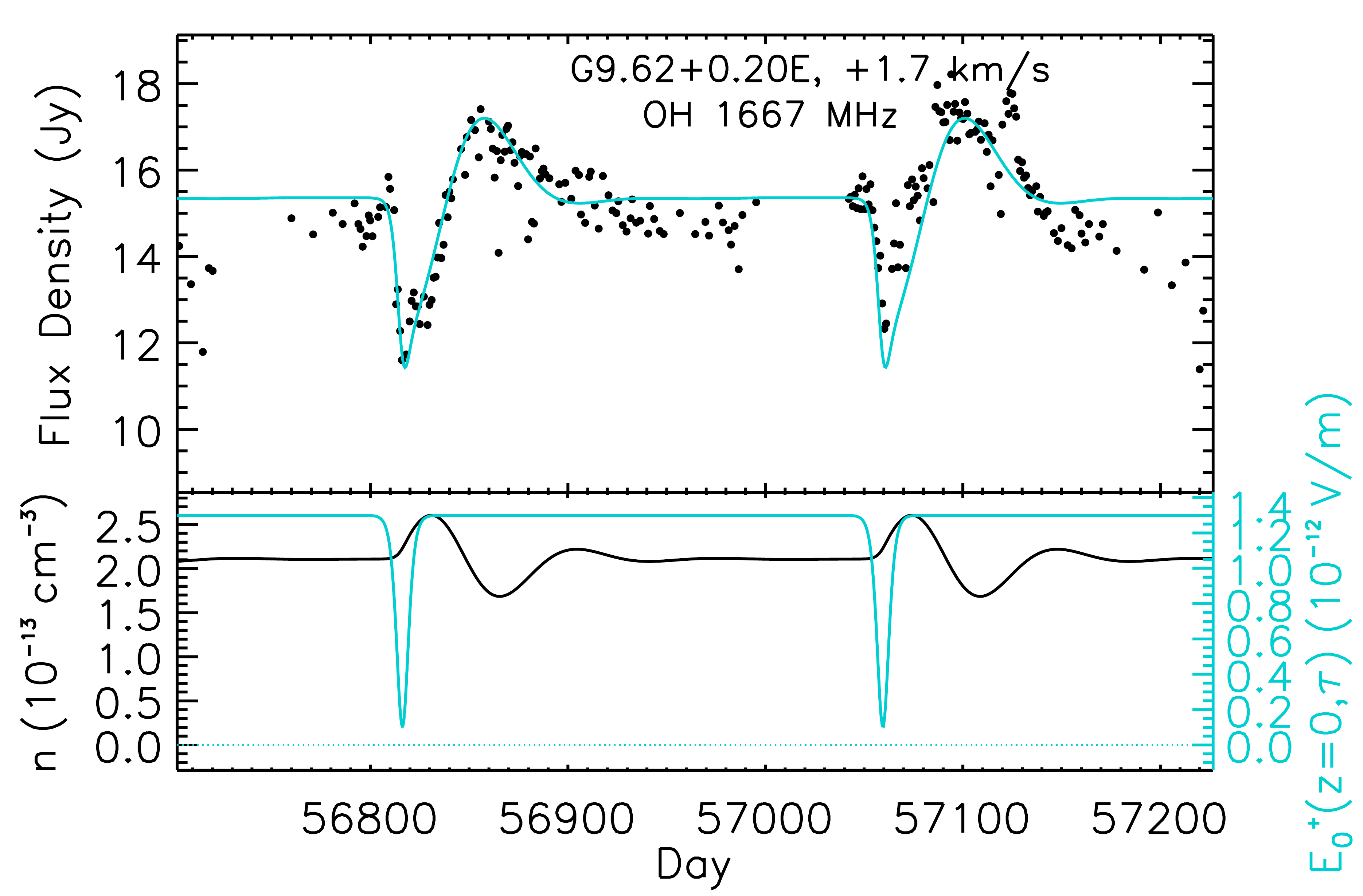}
    \caption{Similar to Figure \ref{fig:OH1667fit} for the OH~1667~MHz feature at $v_{\mathrm{lsr}} = +1.7$~km s$^{-1}$ but with a seed electric field to initiate the periodic flaring. For this system the pump signal is set to a constant level $\Lambda_{\mathrm{0}} = 4.2 \times 10^{-19}$~cm$^{-3}$s$^{-1}$ to achieve an initial population inversion. The seed signal $E^+_0\left(\tau\right)$ at the input of a single sample (bottom panel; cyan curve) has the same form as that given in equation (\ref{eq:pump}) but with a constant electric field of $1.3\times 10^{-12}$~V~m$^{-1}$ and series of pulses of $-1.2\times 10^{-12}$~V~m$^{-1}$ in amplitude (i.e., for the first flare; note the negative sign), $T_1 = 1.8 \times 10^7$~s and $T_2 = 1.2 \times 10^6$~s, while $T_\mathrm{p} = 3.5 \times 10^5$~s and $L = 7.5 \times 10^{14}$~cm, as always.}
    \label{fig:OH1667fit-seed}
\end{figure}

\begin{figure}
    \centering
    \includegraphics[width=0.47\textwidth]{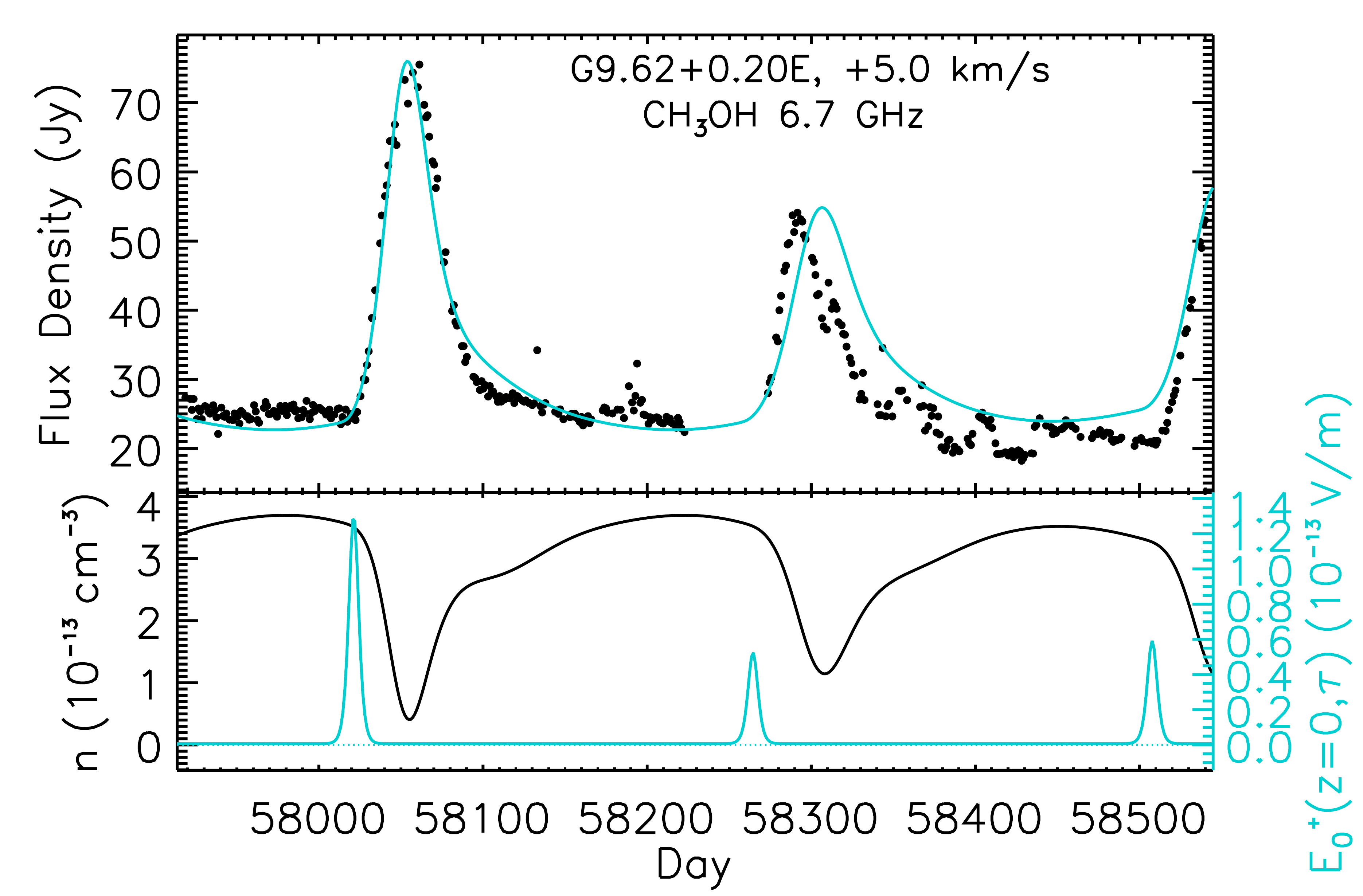}
    \caption{Similar to Figure \ref{fig:OH1667fit-seed} but for the methanol 6.7~GHz feature at $v_{\mathrm{lsr}} = +5$~km s$^{-1}$ with a seed electric field to initiate the periodic flaring. For this system the pump signal is set to a constant level $\Lambda_{\mathrm{0}} = 4.0 \times 10^{-19}$~cm$^{-3}$s$^{-1}$ to achieve an initial population inversion. The seed signal $E^+_0\left(\tau\right)$ at the input of a single sample (bottom panel; cyan curve) has a constant electric field of $7.0\times 10^{-16}$~V~m$^{-1}$ and series of pulses of $1.3\times 10^{-13}$~V~m$^{-1}$ in amplitude (i.e., for the first flare), $T_1 = 1.9 \times 10^7$~s and $T_2 = 1.1 \times 10^6$~s, while $T_\mathrm{p} = 3.5 \times 10^5$~s and $L = 7.5 \times 10^{14}$~cm, as always. This seed field model cannot simultaneously match successive flares nearly as well as the pump excitation realization of Figure \ref{fig:Methanol6.7-5kmfit} for the same feature.}
    \label{fig:6.7GHzfit-seed}
\end{figure}

Another approach to model flares using the MBE (equations \ref{eq:dN/dt} to \ref{eq:dE/dz}) is to use a seed field instead of a pump excitation. That would imply setting $\Lambda_{\mathrm{n}}$ to a constant value in equation (\ref{eq:dN/dt}) to achieve the initial population inversion and adding a trigger or seed as an input to the electric field in equation (\ref{eq:dE/dz}) to model the observed flux densities. For G9.62+0.20E the seed field would have to be periodic and one can use a form similar to that for the pump signal in equation (\ref{eq:pump}), but with $\Lambda_{0}$ and $\Lambda_{1,\mathrm{m}}$ replaced with a constant and pulsed electric field amplitude, respectively. We have actually attempted this approach for different transitions, sometimes with good success. A particularly good example is shown in Figure \ref{fig:OH1667fit-seed} for the same OH~1667~MHz $v_{\mathrm{lsr}} = +1.7$~km s$^{-1}$ feature previously discussed in Sec. \ref{sec:unusual} and presented in Figure \ref{fig:OH1667fit}. However, not all fits fared as well. For instance, we show in Figure \ref{fig:6.7GHzfit-seed} an attempt at modelling the methanol 6.7~GHz $v_{\mathrm{lsr}} = +5$~km s$^{-1}$ feature of Figure \ref{fig:Methanol6.7-5kmfit} in Sec. \ref{sec:6.7GHz}. A comparison between the two results reveals that the seed field excitation model could not simultaneously match successive flares nearly as well as the pump excitation realization could. Still, based on the quality of the OH~1667~MHz model of Figure \ref{fig:OH1667fit-seed} alone we must allow for the possibility that flaring in G9.62+0.20E could also result from triggering due to an ambient seed electric field. However, one would then have to accept relatively low gains in signals from the input to the output for some of our maser/superradiance systems. For the example shown in Figure \ref{fig:OH1667fit-seed} the gain in the electric field is only on the order of ten. 

Our models yield estimates of parameters that provide important information about the environmental conditions of the observed sources. These include the population inversion column density $nL$, as well as the non-coherent relaxation and dephasing time-scales $T_1$ and $T_2$, respectively. As mentioned earlier, we find only slight variations in these time-scales among the four transitions analysed. The value of $T_1$, which can be mainly associated to inelastic collisions, is on the order of several months (220 days) while $T_2$, for elastic collisions, is found to be on the order of several days (12.6~d). The elastic collision time-scale can be estimated with $\tau_{\mathrm{coll}}\approx 1/\left(\sigma_{\mathrm{g}}\bar{v}n_{{\mathrm{H}}_2}\right)$, where $\sigma_{\mathrm{g}}$ and $n_{{\mathrm{H}}_2}$ are the geometrical cross-sectional area and density of H$_2$, the main collision partner of the OH and methanol molecules, and $\bar{v}$ the mean relative velocity between colliding partners. For example, we have $\sigma_{\mathrm{g}} \approx 4\times 10 ^{-16}$~cm$^2$ for OH--H$_2$ collisions \citep{Offer1994} and by setting $\tau_{\mathrm{coll}} = T_2 =  1.1 \times 10^6$~s and $\bar{v} \sim 1$~km~s$^{-1}$ (corresponding to $T \approx 100$~K), we find $n_{{\mathrm{H}}_2} \approx 2 \times 10^ 4$~cm$^{-3}$. This figure is on the lower end but is consistent with expected ranges for OH and methanol. In particular, the methanol~6.7~GHz line is believed to be radiatively pumped for $10^4$~cm$^{-3}\leq n_{{\mathrm{H}}_2}\leq 10^{9}$~cm$^{-3}$ and kinetic temperatures $T_{\mathrm{k}}<T_{\mathrm{d}}$, with $T_{\mathrm{d}}$ the dust temperature \citep{Cragg2002, Cragg2005}.

However, still according to the calculations of \citet{Cragg2005} a gas density of $10^4\,\mathrm{cm}^{-3}$ could be problematic for methanol in that it would require unrealistically long sample length to maintain a population inversion with the expected specific column density and line width (see their Fig. 4). But we note that a relatively small change in our evaluation of the gas density could bring us in the plausible conditions range. For example, we directly associate the value obtained for $T_2$ to the (inverse of the) collision rate which, strictly speaking, is unlikely to be correct. As previously mentioned, $T_2$ is related to non-coherent dephasing between the two maser/superradiance energy levels within our MBE framework. It is thus likely that the true collision rate is higher than we calculate, with a corresponding time-scale shorter than our estimate for $T_2$ and leading to higher gas densities. Furthemore, we should also keep in mind that the use of more complete models in the future (i.e., three-dimensional, see below, and/or finite-bandwidth MBE) might to some extent alter the value for some parameters entering our models. The estimates for the $T_1$ and $T_2$ time-scales may be refined as our study of superradiance progresses.

Finally, it is important to address the effect the presence of superradiance signal may have on the characteristics of the observed radiation. Superradiance is an intrinsically coherent phenomenon, as can be attested from the fact that the intensity of corresponding bursts of radiation scales with the square of the number of molecules involved while their temporal duration is inversely proportional to that number \citep{Dicke1954,Feld1980,Gross1982,Rajabi2020}. Despite this singular characteristic, one should not automatically expect that radiation detected from astronomical sources exhibiting superradiance flares to bear the imprint coherence. To understand this one must consider the size of a single superradiance sample in relation to that of the emitting region it is embedded in. For example, we find that the OH~1665~MHz sample obtained for the fit presented in Figure \ref{fig:OH1665fit} has a radius limited to $\sim 10^{-5}$~au to ensure a Fresnel number of unity, as phase coherence could not be maintained over a sample wider than this. This implies that a maser-hosting region with a spot size of, say, 10~au undergoing a superradiance event would necessarily break up in a very large number of independent and uncorrelated samples, which would fan out over the exit solid angle set by the geometry of hosting region \citep{Gross1982}. Although each individual sample radiates coherently, the total intensity from the group of independent samples will appear non-coherent to an observer. We find that our individual OH sample radiates with a peak superradiance flux density of $\approx 5\times 10^{-36}$~erg~s$^{-1}$~cm$^{-2}$ at 5.2~kpc (or $\sim 10^{-5}$~Jy in the bandwidth of $\sim 10^{-7}$~Hz stemming from the duration of a flare), which implies that only a small fraction of a typical maser-hosting volume of gas would need to undergo a flaring episode in order to match the observed flux densities.  

Incidentally, this discussion highlights a further limitation of the one-dimensional MBE framework used in this paper. That is, these simple calculations imply the existence of transverse variations in population inversion levels (i.e., in its column density) across the spot size of real astronomical systems, as could be intuitively expected. These variations coupled with the critical threshold required for attaining the superradiance regime will lead to the formation of independent flaring regions of differing intensities and time-scales (and non-flaring  samples in some regions); see equation (\ref{eq:TR}) and \citet{Rajabi2020}. This could result, for example, in a decrease in apparent spot size and radiation beaming during a flaring episode, similar to what is predicted for masers \citep{Elitzur1992b}. We thus expect that in more realistic situations the superradiance problem is inherently three-dimensional in nature. The development of a three-dimensional MBE framework \citep{Gross1982} is therefore required for obtaining a better understanding of superradiance in astronomical media.

\section*{Acknowledgements}
M.H. dedicates this work to the memory of his friend and mentor Thomas G. Phillips. M.H.'s research is funded through the Natural Sciences and Engineering Research Council of Canada Discovery Grant RGPIN-2016-04460. F.R.'s research at Perimeter Institute was supported in part by the Government of Canada through the Department of Innovation, Science and Economic Development Canada and by the Province of Ontario through the Ministry of Economic Development, Job Creation and Trade. The 6.7 GHz methanol maser data obtained by the Hitachi 32-m telescope is a part of the Ibaraki 6.7 GHz Methanol Maser Monitor (iMet) program. The iMet program is partially supported by the Inter-university collaborative project `Japanese VLBI Network (JVN)’ of NAOJ and JSPS KAKENHI grant no. JP24340034, JP21H01120, and JP21H00032 (YY).

\section*{Data availability}
The data underlying this article will be shared on reasonable request to the corresponding author.

\bibliographystyle{mnras}
\bibliography{scibib}


\label{lastpage}

\end{document}